\documentclass[12pt]{article}

\newcommand{\mrm}[1]{\mathrm{#1}}
 
\newcommand{\alphas}{\alpha_{\mathrm{s}}}

\newcommand{\kT}{k_{\perp}}
\newcommand{\pT}{p_{\perp}}
\newcommand{\lessim}{\raisebox{-0.8mm}%
{\hspace{1mm}$\stackrel{<}{\sim}$\hspace{1mm}}}
\newcommand{\gtrsim}{\raisebox{-0.8mm}%
{\hspace{1mm}$\stackrel{>}{\sim}$\hspace{1mm}}}
 
\renewcommand{\b}{{\mathrm b}}
\newcommand{\crm}{{\mathrm c}}
\renewcommand{\d}{{\mathrm d}}
\newcommand{\e}{{\mathrm e}}
\newcommand{\f}{{\mathrm f}}
\newcommand{\g}{{\mathrm g}}
\newcommand{\hrm}{{\mathrm h}}

\newcommand{\p}{{\mathrm p}}
\newcommand{\q}{{\mathrm q}}
\newcommand{\s}{{\mathrm s}}
\renewcommand{\t}{{\mathrm t}}
\renewcommand{\u}{{\mathrm u}}
\newcommand{\A}{{\mathrm A}}
\newcommand{\B}{{\mathrm B}}
\newcommand{\D}{{\mathrm D}}
\newcommand{\F}{{\mathrm F}}
\renewcommand{\H}{{\mathrm H}}
\newcommand{\J}{{\mathrm J}}

\renewcommand{\L}{{\mathrm L}}
\newcommand{\M}{{\mathrm M}}
\newcommand{\Q}{{\mathrm Q}}
\newcommand{\R}{{\mathrm R}}

\newcommand{\W}{{\mathrm W}}
\newcommand{\Z}{{\mathrm Z}}
\newcommand{\bbar}{\overline{\mathrm b}}
\newcommand{\cbar}{\overline{\mathrm c}}

\newcommand{\fbar}{\overline{\mathrm f}}

\newcommand{\qbar}{\overline{\mathrm q}}

\newcommand{\tbar}{\overline{\mathrm t}}

\newcommand{\Dbar}{\overline{\mathrm D}}
\newcommand{\Fbar}{\overline{\mathrm F}}
\newcommand{\Qbar}{\overline{\mathrm Q}}

\newcommand{\sq}{\tilde{\mathrm q}}
\newcommand{\sqs}{\tilde{\mathrm q}^*}
\newcommand{\sqbar}{\overline{\tilde{\mathrm{q}}}}
\newcommand{\sg}{\tilde{\mathrm{g}}}
\newcommand{\tp}{\tilde{\mathrm t}}
\newcommand{\tm}{\tilde{\mathrm t}^*}
\newcommand{\sd}{\tilde{\mathrm d}}
\newcommand{\su}{\tilde{\mathrm u}}
\newcommand{\sch}{\tilde{\mathrm c}}
\newcommand{\sst}{\tilde{\mathrm s}}
\newcommand{\st}{\tilde{\mathrm t}}
\newcommand{\sbo}{\tilde{\mathrm b}}
\newcommand{\sbs}{\tilde{\mathrm b}^*}
\newcommand{\se}{\tilde{\mathrm e}}
\newcommand{\smu}{\tilde{\mu}}
\newcommand{\stau}{\tilde{\tau}}
\newcommand{\snu}{\tilde{\nu}}
\newcommand{\sell}{\tilde{\ell}}

\newcommand{\glu}{\tilde{\mathrm g}}
\newcommand{\chio}{\tilde{\chi}}
\newcommand{\chip}{\tilde{\chi}^{\pm}}
\newcommand{\chim}{\tilde{\chi}^{\mp}}
\newcommand{\grav}{\tilde{\mathrm G}}

\newcommand{\ee}{\e^+\e^-}
\newcommand{\ep}{\e\p}
\newcommand{\pp}{\p\p}

\newcommand{\gammaZ}{\gamma^* / \Z^0}
\newcommand{\gast}{\gamma^*}
 
\newenvironment{Itemize}{\begin{list}{$\bullet$}%
{\setlength{\topsep}{0.2mm}\setlength{\partopsep}{0.2mm}%
\setlength{\itemsep}{0.2mm}\setlength{\parsep}{0.2mm}}}%
{\end{list}}
\newcounter{enumct}
\newenvironment{Enumerate}{\begin{list}{\arabic{enumct}.}%
{\usecounter{enumct}\setlength{\topsep}{0.2mm}%
\setlength{\partopsep}{0.2mm}\setlength{\itemsep}{0.2mm}%
\setlength{\parsep}{0.2mm}}}{\end{list}}

\begin{document}
 
\sloppy
 
\pagestyle{empty}
 
\begin{flushright}
LU TP 00--30\\
hep-ph/0010017\\
October 2000
\end{flushright}
 
\vspace{\fill}
 
\begin{center}
{\LARGE\bf High-Energy-Physics Event Generation with}\\[5mm]
{\Huge\bf P}{\LARGE\bf YTHIA} \  {\Huge\bf 6.1}   \\[20mm]
{\Large Torbj\"orn Sj\"ostrand$^1$, Patrik Ed\'en$^2$, 
Christer Friberg$^1$,} \\[1mm] 
{\Large  Leif L\"onnblad$^1$, Gabriela Miu$^1$, 
Stephen Mrenna$^3$} \\[1mm] 
{\Large and Emanuel Norrbin$^1$} \\[8mm]
{\large 1) Department of Theoretical Physics, Lund University,\\
S\"olvegatan 14A, S-223 62 Lund, Sweden} \\[2mm]
{\large 2) NORDITA, Blegdamsvej 17, DK-2100 Copenhagen, Denmark}\\[2mm]
{\large 3) Physics Department, University of California at Davis,\\
One Shields Avenue, Davis, CA 95616, USA}\\
\end{center}

\vspace{\fill}
 
\begin{center}
\bf{Abstract} 
\end{center}
\vspace{-0.5\baselineskip}
\noindent
\textsc{Pythia} version 6 represents a merger of the 
\textsc{Pythia}~5, \textsc{Jetset}~7 and \textsc{SPythia} 
programs, with many improvements. It can be used to generate 
high-energy-physics `events', i.e. sets of outgoing particles 
produced in the interactions between two incoming particles. The 
objective is to provide as accurate as possible a representation
of event properties in a wide range of reactions. The underlying 
physics is not understood well enough to give an exact description; 
the programs therefore contain a combination of analytical results 
and various models. The emphasis in this article is on new aspects, 
but a few words of general introduction are included. Further 
documentation is available on the web.

\vspace{\fill}

\noindent
CPC subject index: 11.2

\vspace{\fill}
 
\clearpage
\pagestyle{plain}
\setcounter{page}{1}

\section*{New Version Summary}

\noindent
{\em Title of program:} \textsc{Pythia}

\noindent
{\em Version number:} 6.154

\vspace{0.2\baselineskip}\noindent
{\em Catalogue identifier:} --- 

\vspace{0.2\baselineskip}\noindent
{\em Distribution format:} uuencoded compressed tar file 

\vspace{0.2\baselineskip}\noindent
{\em References to most recent previous versions:}
Computer Physics Communications {\bf 82} (1994) 74 and
Computer Physics Communications {\bf 101} (1997) 232, respectively

\vspace{0.2\baselineskip}\noindent
{\em Catalogue identifiers of most recent previous versions:}
ACTU

\vspace{0.2\baselineskip}\noindent
{\em Authors of most recent previous versions:}
T. Sj\"ostrand, and S. Mrenna, respectively

\vspace{0.2\baselineskip}\noindent
{\em Does the new version supersede the previous versions?:} yes 

\vspace{0.2\baselineskip}\noindent
{\em Computers for which the new program is designed and others on which
it has been tested:}\\
{\em Computer:} DELL Precision 210 and any other machine with 
a Fortran 77 compiler \\
{\em Installation:} Lund University \\
{\em Operating system under which the new program has
been tested:} Red Hat Linux 6.2

\vspace{0.2\baselineskip}\noindent
{\em Programming language used:} Fortran 77;
is also fully compatible with Fortran 90, i.e. does not make
use of any obsolescent features of the Fortran 90 standard

\vspace{0.2\baselineskip}\noindent
{\em Memory required to execute with typical data:} about 800 kwords

\vspace{0.2\baselineskip}\noindent
{\em No.\ of bits in word:} 32 (double precision real uses two words)

\vspace{0.2\baselineskip}\noindent
{\em No.\ of processors used:} 1

\vspace{0.2\baselineskip}\noindent
{\em Has the code been vectorized or parallelized?:} no

\vspace{0.2\baselineskip}\noindent
{\em No.\ of bytes in distributed program:} about 1.8 Mb

\vspace{0.2\baselineskip}\noindent
{\em Keywords:} QCD, standard model, beyond standard model,
hard scattering, $\ee$ annihilation, leptoproduction, photoproduction,
hadronic processes, high-$\pT$ scattering, prompt photons,
gauge bosons, Higgs physics, parton distribution functions, 
jet production, parton showers, fragmentation, hadronization, 
beam remnants, multiple interactions, particle decays, 
event measures 

\vspace{0.2\baselineskip}\noindent
{\em Nature of physical problem:} high-energy collisions between 
elementary particles normally give rise to complex final states,
with large multiplicities of hadrons, leptons, neutrinos and photons.
The relation between these final states and the underlying 
physics description is not a simple one, for two main reasons. 
Firstly, we do not even in principle have a complete understanding 
of the physics. Secondly, any analytical approach is made 
intractable by the large multiplicities.  

\vspace{0.2\baselineskip}\noindent
{\em Method of solution:} complete events are generated by Monte Carlo 
methods. The complexity is mastered by a subdivision of the full 
problem into a set of simpler separate tasks.
All main aspects of the events are simulated, such as
hard-process selection, initial- and final-state radiation, beam
remnants, fragmentation, decays, and so on. Therefore events should be
directly comparable with experimentally observable ones. The programs
can be used to extract physics from comparisons with existing
data, or to study physics at future experiments.

\vspace{0.2\baselineskip}\noindent
{\em Restrictions on the complexity of the problem:} depends on the 
problem studied

\vspace{0.2\baselineskip}\noindent
{\em Typical running time:} 10--1000 events per second, depending on
process studied

\vspace{0.2\baselineskip}\noindent
{\em Unusual features of the program:} none
 
\clearpage

\section{Introduction}

The \textsc{Pythia} and \textsc{Jetset} programs \cite{PYTSET} are 
frequently used for 
event generation in high-energy physics. The emphasis is on 
multiparticle production in collisions between elementary particles. 
This in particular means hard interactions in $\ee$, $\pp$ and $\ep$ 
colliders, although also other applications are envisaged. The 
programs can be used both to compare with existing data, for 
physics studies or detector corrections, and to explore possibilities
at present or future experiments. The programs are intended to generate 
complete events, in as much detail as experimentally observable ones, 
within the bounds of our current understanding of the underlying 
physics. The quantum mechanical variability between events in
nature is here replaced by Monte Carlo methods, to obtain `correctly' 
both the average behaviour and the amount of fluctuations. 
Many of the components of the programs represent original research, 
in the sense that models have been developed and implemented for a 
number of aspects not covered by standard theory. 

Although originally conceived separately, the \textsc{Pythia} 
\cite{PYTHIA} and \textsc{Jetset} \cite{JETSET} programs today are 
so often used together that they have here been joined under the 
\textsc{Pythia} header. To this has been added the code of 
\textsc{SPythia} \cite{SPYTHIA}, an extension of \textsc{Pythia} that
also covers the generation of supersymmetric processes. The current 
article is not intended to give a complete survey of all the program
elements or all the physics --- we refer to the long manual and further
documentation on the \textsc{Pythia} web page\\
\hspace*{\fill}
\texttt{http://www.thep.lu.se/}$\sim$\texttt{torbjorn/Pythia.html}
\hspace*{\fill}\\
for this. Instead the objective is to give a survey of new features 
since the latest official publications, with some minimal amount of
background material to tie the story together. Many of the advances are 
related to physics studies, which are further described in separate
articles. Others are of a more technical nature, or have been of too
limited a scope to result in individual publications (so far).

Some of the main topics are:
\begin{Itemize}
\item An improved simulation of supersymmetric physics, with several
new processes.
\item Many new processes of beyond-the-standard-model physics, in
areas such as technicolor and doubly-charged Higgses.
\item An expanded description of QCD processes in virtual-photon
interactions, combined with a new machinery for the flux of virtual 
photons from leptons.
\item Initial-state parton showers are matched to the next-to-leading
order matrix elements for gauge boson production. 
\item Final-state parton showers are matched to a number of different
first-order matrix elements for gluon emission, including full
mass dependence.
\item The hadronization description of low-mass strings has been
improved, with consequences especially for heavy-flavour production.
\item An alternative baryon production model has been introduced.
\item Colour rearrangement is included as a new option, and several
alternative Bose-Einstein descriptions are added.  
\end{Itemize} 
Many further examples will be given. In the process, the total size of 
the program code has almost doubled in the six years since the previous 
main publication.

The report is subdivided so that the physics news are highlighted
in section 2 and the programming ones (plus a few more physics ones) 
in section 3. Section 4 contains some concluding remarks and an outlook.
 
\section{Physics News}
 
For the description of a typical high-energy event, an event
generator should contain a simulation of several physics aspects.
If we try to follow the evolution of an event in some semblance of
a time order, one may arrange these aspects as follows:
\begin{Enumerate}
\item Initially two beam particles are coming in towards each other.
      Normally each particle is characterized by a set of parton 
      distributions, which defines the partonic substructure in terms 
      of flavour composition and energy sharing.
\item One shower initiator parton from each beam starts off
      a sequence of branchings, such as $\q \to \q \g$, which build up
      an initial-state shower.
\item One incoming parton from each of the two showers
      enters the hard process, where then a number of
      outgoing partons are produced, usually two.
      It is the nature of this process that determines the main
      characteristics of the event.
\item The hard process may produce a set of short-lived resonances,
      like the $\Z^0/\W^{\pm}$ gauge bosons, whose decay to normal 
      partons has to be considered in close association with the 
      hard process itself.
\item The outgoing partons may branch as well, to build up
      final-state showers.
\item In addition to the hard process considered above, further
      semihard interactions may occur between the other partons 
      of two incoming hadrons.
\item When a shower initiator is taken out of a beam particle,
      a beam remnant is left behind. This remnant may have
      an internal structure, and a net colour charge that relates
      it to the rest of the final state.
\item The QCD confinement mechanism ensures that the outgoing quarks 
      and gluons are not observable, but instead fragment to colour 
      neutral hadrons.
\item Normally the fragmentation mechanism can be seen as occurring
      in a set of separate colour singlet subsystems, but 
      interconnection effects such as colour rearrangement or 
      Bose--Einstein may complicate the picture.
\item Many of the produced hadrons are unstable and decay further.
\end{Enumerate}
 
In the following subsections, we will survey updates of the above 
aspects, not in the same order as given here, but rather in the order 
in which they appear in the program execution, i.e. starting with the 
hard process.
 
\subsection{Physics Subprocesses}

\begin{table}[t]
\caption{Subprocesses, part 1: standard model, according to the 
subprocess numbering of \textsc{Pythia}.`$\f$' denotes a fermion 
(quark or lepton), `$\Q$' a heavy quark and `$\F$' a heavy fermion.  
\protect\label{table1} } 
\begin{tabular}[t]{|rl|@{\protect\rule[-2mm]{0mm}{6mm}}}
\hline
No. & Subprocess  \\ 
\hline
\multicolumn{2}{|l|@{\protect\rule[-2mm]{0mm}{7mm}}}{Hard QCD processes:} \\
 11 & $\f_i \f_j \to \f_i \f_j$ \\
 12 & $\f_i \fbar_i \to \f_k \fbar_k$ \\
 13 & $\f_i \fbar_i \to \g \g$  \\
 28 & $\f_i \g \to \f_i \g$   \\
 53 & $\g \g \to \f_k \fbar_k$  \\
 68 & $\g \g \to \g \g$  \\ 
\hline
\multicolumn{2}{|l|@{\protect\rule[-2mm]{0mm}{7mm}}}{Soft QCD processes:} \\
 91 & elastic scattering  \\
 92 & single diffraction ($XB$)  \\
 93 & single diffraction ($AX$)  \\
 94 & double diffraction  \\
 95 & low-$p_{\perp}$ production   \\ 
\hline
\multicolumn{2}{|l|@{\protect\rule[-2mm]{0mm}{7mm}}}{Open heavy flavour:} \\
\multicolumn{2}{|l|@{\protect\rule[-2mm]{0mm}{6mm}}}{(also fourth generation)} \\
 81 & $\f_i \fbar_i \to \Q_k \Qbar_k$  \\
 82 & $\g \g \to \Q_k \Qbar_k$  \\
 83 & $\q_i \f_j \to \Q_k \f_l$  \\
 84 & $\g \gamma \to \Q_k \Qbar_k$  \\
 85 & $\gamma \gamma \to \F_k \Fbar_k$  \\
\hline
\multicolumn{2}{|l|@{\protect\rule[-2mm]{0mm}{7mm}}}{Closed heavy flavour:} \\
 86 & $\g \g \to \J/\psi \g$  \\
 87 & $\g \g \to \chi_{0\crm} \g$   \\
 88 & $\g \g \to \chi_{1\crm} \g$  \\
 89 & $\g \g \to \chi_{2\crm} \g$   \\
104 & $\g \g \to \chi_{0\crm}$ \\   
105 & $\g \g \to \chi_{2\crm}$ \\ 
106 & $\g \g \to \J/\psi \gamma$  \\
107 & $\g \gamma \to \J/\psi \g$  \\
108 & $\gamma \gamma \to \J/\psi \gamma$  \\
\hline
\end{tabular}
\hfill
\begin{tabular}[t]{|rl|@{\protect\rule[-2mm]{0mm}{6mm}}}
\hline
No. & Subprocess \\ 
\hline
\multicolumn{2}{|l|@{\protect\rule[-2mm]{0mm}{7mm}}}{$\W / \Z$ production:} \\
  1 & $\f_i \fbar_i \to \gammaZ$  \\
  2 & $\f_i \fbar_j \to \W^{\pm}$  \\
 22 & $\f_i \fbar_i \to \Z^0 \Z^0$   \\
 23 & $\f_i \fbar_j \to \Z^0 \W^{\pm}$   \\
 25 & $\f_i \fbar_i \to \W^+ \W^-$  \\
 15 & $\f_i \fbar_i \to \g \Z^0$  \\
 16 & $\f_i \fbar_j \to \g \W^{\pm}$   \\
 30 & $\f_i \g \to \f_i \Z^0$  \\
 31 & $\f_i \g \to \f_k \W^{\pm}$   \\
 19 & $\f_i \fbar_i \to \gamma \Z^0$   \\
 20 & $\f_i \fbar_j \to \gamma \W^{\pm}$   \\
 35 & $\f_i \gamma \to \f_i \Z^0$   \\
 36 & $\f_i \gamma \to \f_k \W^{\pm}$  \\
 69 & $\gamma \gamma \to \W^+ \W^-$  \\
 70 & $\gamma \W^{\pm} \to \Z^0 \W^{\pm}$  \\
\hline 
\multicolumn{2}{|l|@{\protect\rule[-2mm]{0mm}{7mm}}}{Prompt photons:} \\
 14 & $\f_i \fbar_i \to \g \gamma$  \\
 18 & $\f_i \fbar_i \to \gamma \gamma$   \\
 29 & $\f_i \g \to \f_i \gamma$   \\
114 & $\g \g \to \gamma \gamma$  \\
115 & $\g \g \to \g \gamma$   \\
\hline
\multicolumn{2}{|l|@{\protect\rule[-2mm]{0mm}{7mm}}}{Deep inelastic scatt.:} \\
 10 & $\f_i \f_j \to \f_i \f_j$  \\
 99 & $\gast \f_i \to f_i$ \\
\hline
\multicolumn{2}{|l|@{\protect\rule[-2mm]{0mm}{7mm}}}{Photon-induced:} \\
 33 & $\f_i \gamma \to \f_i \g$  \\
 34 & $\f_i \gamma \to \f_i \gamma$   \\
 54 & $\g \gamma \to \f_k \fbar_k$  \\
 58 & $\gamma \gamma \to \f_k \fbar_k$   \\
\hline
\end{tabular}
\hfill
\begin{tabular}[t]{|rl|@{\protect\rule[-2mm]{0mm}{6mm}}}
\hline
No. & Subprocess \\ 
\hline
131 & $\f_i \gast_{\mrm{T}} \to \f_i \g$ \\
132 & $\f_i \gast_{\mrm{L}} \to \f_i \g$ \\
133 & $\f_i \gast_{\mrm{T}} \to \f_i \gamma$ \\
134 & $\f_i \gast_{\mrm{L}} \to \f_i \gamma$ \\
135 & $\g \gast_{\mrm{T}} \to \f_i \fbar_i$ \\
136 & $\g \gast_{\mrm{L}} \to \f_i \fbar_i$ \\
137 & $\gast_{\mrm{T}} \gast_{\mrm{T}} \to \f_i \fbar_i$ \\
138 & $\gast_{\mrm{T}} \gast_{\mrm{L}} \to \f_i \fbar_i$ \\
139 & $\gast_{\mrm{L}} \gast_{\mrm{T}} \to \f_i \fbar_i$ \\
140 & $\gast_{\mrm{L}} \gast_{\mrm{L}} \to \f_i \fbar_i$ \\
 80 & $\q_i \gamma \to \q_k \pi^{\pm}$ \\
\hline
\multicolumn{2}{|l|@{\protect\rule[-2mm]{0mm}{7mm}}}{Light SM Higgs:} \\
  3 & $\f_i \fbar_i \to \hrm^0$  \\
 24 & $\f_i \fbar_i \to \Z^0 \hrm^0$  \\
 26 & $\f_i \fbar_j \to \W^{\pm} \hrm^0$  \\
102 & $\g \g \to \hrm^0$   \\
103 & $\gamma \gamma \to \hrm^0$  \\
110 & $\f_i \fbar_i \to \gamma \hrm^0$  \\
121 & $\g \g \to \Q_k \Qbar_k \hrm^0$  \\
122 & $\q_i \qbar_i \to \Q_k \Qbar_k \hrm^0$  \\
123 & $\f_i \f_j \to \f_i \f_j \hrm^0$  \\
124 & $\f_i \f_j \to \f_k \f_l \hrm^0$  \\
\hline
\multicolumn{2}{|l|@{\protect\rule[-2mm]{0mm}{7mm}}}{Heavy SM Higgs:} \\
  5 & $\Z^0 \Z^0 \to \hrm^0$  \\
  8 & $\W^+ \W^- \to \hrm^0$  \\
 71 & $\Z^0_{\mrm{L}} \Z^0_{\mrm{L}} \to \Z^0_{\mrm{L}} 
\Z^0_{\mrm{L}}$ \\
 72 & $\Z^0_{\mrm{L}} \Z^0_{\mrm{L}} \to \W^+_{\mrm{L}} 
\W^-_{\mrm{L}}$ \\
 73 & $\Z^0_{\mrm{L}} \W^{\pm}_{\mrm{L}} \to \Z^0_{\mrm{L}} 
\W^{\pm}_{\mrm{L}}$  \\
 76 & $\W^+_{\mrm{L}} \W^-_{\mrm{L}} \to \Z^0_{\mrm{L}} 
\Z^0_{\mrm{L}}$  \\
 77 & $\W^{\pm}_{\mrm{L}} \W^{\pm}_{\mrm{L}} \to 
\W^{\pm}_{\mrm{L}} \W^{\pm}_{\mrm{L}}$  \\
\hline
\end{tabular}
\end{table}

\begin{table}[t]
\caption{Subprocesses, part 2: beyond the standard model non-SUSY, 
with notation as above.
\protect\label{table2} } 
\begin{tabular}[t]{|rl|@{\protect\rule[-2mm]{0mm}{6mm}}}
\hline
No. & Subprocess \\ 
\hline
\multicolumn{2}{|l|@{\protect\rule[-2mm]{0mm}{7mm}}}{BSM Neutral Higgses:} \\
151 & $\f_i \fbar_i \to \H^0$ \\
152 & $\g \g \to \H^0$  \\
153 & $\gamma \gamma \to \H^0$  \\
171 & $\f_i \fbar_i \to \Z^0 \H^0$  \\
172 & $\f_i \fbar_j \to \W^{\pm} \H^0$  \\
173 & $\f_i \f_j \to \f_i \f_j \H^0$   \\
174 & $\f_i \f_j \to \f_k \f_l \H^0$  \\
181 & $\g \g \to \Q_k \Qbar_k \H^0$  \\
182 & $\q_i \qbar_i \to \Q_k \Qbar_k \H^0$   \\
156 & $\f_i \fbar_i \to \A^0$  \\
157 & $\g \g \to \A^0$  \\
158 & $\gamma \gamma \to \A^0$  \\
176 & $\f_i \fbar_i \to \Z^0 \A^0$   \\
177 & $\f_i \fbar_j \to \W^{\pm} \A^0$  \\
178 & $\f_i \f_j \to \f_i \f_j \A^0$  \\
179 & $\f_i \f_j \to \f_k \f_l \A^0$   \\
186 & $\g \g \to \Q_k \Qbar_k \A^0$ \\
187 & $\q_i \qbar_i \to \Q_k \Qbar_k \A^0$  \\
\hline
\multicolumn{2}{|l|@{\protect\rule[-2mm]{0mm}{7mm}}}{Charged Higgs:} \\
143 & $\f_i \fbar_j \to \H^+$  \\
161 & $\f_i \g \to \f_k \H^+$  \\
\hline
\multicolumn{2}{|l|@{\protect\rule[-2mm]{0mm}{7mm}}}{Higgs pairs:} \\
297 & $\f_i \fbar_j \to \H^{\pm} \hrm^0$ \\ 
298 & $\f_i \fbar_j \to \H^{\pm} \H^0$ \\ 
299 & $\f_i \fbar_i \to \A^0 \hrm^0$ \\ 
300 & $\f_i \fbar_i \to \A^0 \H^0$ \\ 
301 & $\f_i \fbar_i \to \H^+ \H^-$ \\ 
\hline
\end{tabular}
\hfill
\begin{tabular}[t]{|rl|@{\protect\rule[-2mm]{0mm}{6mm}}}
\hline
No. & Subprocess \\ 
\hline
\multicolumn{2}{|l|@{\protect\rule[-2mm]{0mm}{7mm}}}{New gauge bosons:} \\
141 & $\f_i \fbar_i \to \gamma/\Z^0/{\Z'}^0$ \\
142 & $\f_i \fbar_j \to {\W'}^+$ \\
144 & $\f_i \fbar_j \to \R$  \\
\hline
\multicolumn{2}{|l|@{\protect\rule[-2mm]{0mm}{7mm}}}{Technicolor:} \\
149 & $\g \g \to \eta_{\mrm{tc}}$   \\
191 & $\f_i \fbar_i \to \rho_{\mrm{tc}}^0$   \\
192 & $\f_i \fbar_j \to \rho_{\mrm{tc}}^+$   \\
193 & $\f_i \fbar_i \to \omega_{\mrm{tc}}^0$   \\
194 & $\f_i \fbar_i \to \f_k \fbar_k$   \\
195 & $\f_i \fbar_j \to \f_k \fbar_l$   \\
361 & $\f_i \fbar_i \to \W^+_{\mrm{L}} \W^-_{\mrm{L}} $  \\
362 & $\f_i \fbar_i \to \W^{\pm}_{\mrm{L}} \pi^{\mp}_{\mrm{tc}}$   \\
363 & $\f_i \fbar_i \to \pi^+_{\mrm{tc}} \pi^-_{\mrm{tc}}$   \\
364 & $\f_i \fbar_i \to \gamma \pi^0_{\mrm{tc}} $   \\
365 & $\f_i \fbar_i \to \gamma {\pi'}^0_{\mrm{tc}} $   \\
366 & $\f_i \fbar_i \to \Z^0 \pi^0_{\mrm{tc}} $   \\
367 & $\f_i \fbar_i \to \Z^0 {\pi'}^0_{\mrm{tc}} $   \\
368 & $\f_i \fbar_i \to \W^{\pm} \pi^{\mp}_{\mrm{tc}}$ \\
370 & $\f_i \fbar_j \to \W^{\pm}_{\mrm{L}} \Z^0_{\mrm{L}} $  \\
371 & $\f_i \fbar_j \to \W^{\pm}_{\mrm{L}} \pi^0_{\mrm{tc}}$   \\
372 & $\f_i \fbar_j \to \pi^{\pm}_{\mrm{tc}} \Z^0_{\mrm{L}} $ \\
373 & $\f_i \fbar_j \to \pi^{\pm}_{\mrm{tc}} \pi^0_{\mrm{tc}} $   \\
374 & $\f_i \fbar_j \to \gamma \pi^{\pm}_{\mrm{tc}} $   \\
375 & $\f_i \fbar_j \to \Z^0 \pi^{\pm}_{\mrm{tc}} $   \\
376 & $\f_i \fbar_j \to \W^{\pm} \pi^0_{\mrm{tc}} $   \\
377 & $\f_i \fbar_j \to \W^{\pm} {\pi'}^0_{\mrm{tc}}$ \\
\hline
\end{tabular}
\hfill
\begin{tabular}[t]{|rl|@{\protect\rule[-2mm]{0mm}{6mm}}}
\hline
No. & Subprocess \\ 
\hline
\multicolumn{2}{|l|@{\protect\rule[-2mm]{0mm}{7mm}}}{Compositeness:} \\
146 & $\e \gamma \to \e^*$  \\
147 & $\d \g \to \d^*$  \\
148 & $\u \g \to \u^*$  \\
167 & $\q_i \q_j \to \d^* \q_k$  \\
168 & $\q_i \q_j \to \u^* \q_k$  \\
169 & $\q_i \qbar_i \to \e^{\pm} \e^{*\mp}$  \\
165 & $\f_i \fbar_i (\to \gast/\Z^0) \to \f_k \fbar_k$  \\
166 & $\f_i \fbar_j (\to \W^{\pm}) \to \f_k \fbar_l$ \\
\hline
\multicolumn{2}{|l|@{\protect\rule[-2mm]{0mm}{7mm}}}{Doubly-charged Higgs:} \\
341 & $\ell_i \ell_j \to \H_L^{\pm\pm}$ \\
342 & $\ell_i \ell_j \to \H_R^{\pm\pm}$ \\
343 & $\ell_i^{\pm} \gamma \to \H_L^{\pm\pm} \e^{\mp}$ \\
344 & $\ell_i^{\pm} \gamma \to \H_R^{\pm\pm} \e^{\mp}$ \\
345 & $\ell_i^{\pm} \gamma \to \H_L^{\pm\pm} \mu^{\mp}$ \\
346 & $\ell_i^{\pm} \gamma \to \H_R^{\pm\pm} \mu^{\mp}$ \\
347 & $\ell_i^{\pm} \gamma \to \H_L^{\pm\pm} \tau^{\mp}$ \\
348 & $\ell_i^{\pm} \gamma \to \H_R^{\pm\pm} \tau^{\mp}$ \\
349 & $\f_i \fbar_i \to \H_L^{++} \H_L^{--}$ \\ 
350 & $\f_i \fbar_i \to \H_R^{++} \H_R^{--}$ \\ 
351 & $\f_i \f_j \to \f_k \f_l \H_L^{\pm\pm}$  \\
352 & $\f_i \f_j \to \f_k \f_l \H_R^{\pm\pm}$  \\
\hline
\multicolumn{2}{|l|@{\protect\rule[-2mm]{0mm}{7mm}}}{Leptoquarks:} \\
145 & $\q_i \ell_j \to \L_{\Q}$  \\
162 & $\q \g \to \ell \L_{\Q}$  \\
163 & $\g \g \to \L_{\Q} \overline{\L}_{\Q}$  \\
164 & $\q_i \qbar_i \to \L_{\Q} \overline{\L}_{\Q}$ \\
\hline
\end{tabular}
\end{table}

\begin{table}[t]
\caption{Subprocesses, part 3: SUSY, with notation as above. 
A trailing $+$ on a final state indicates that the charge-conjugated 
one is included as well.
\protect\label{table3} } 
\begin{tabular}[t]{|rl|@{\protect\rule[-2mm]{0mm}{6mm}}}
\hline
No. & Subprocess \\ 
\hline
\multicolumn{2}{|l|@{\protect\rule[-2mm]{0mm}{7mm}}}{SUSY:} \\
201 & $\f_i \fbar_i \to \se_L \se_L^*$  \\
202 & $\f_i \fbar_i \to \se_R \se_R^*$  \\
203 & $\f_i \fbar_i \to \se_L \se_R^* +$ \\
204 & $\f_i \fbar_i \to \smu_L \smu_L^*$ \\
205 & $\f_i \fbar_i \to \smu_R \smu_R^*$ \\
206 & $\f_i \fbar_i\to\smu_L \smu_R^* +$ \\
207 & $\f_i \fbar_i\to\stau_1 \stau_1^*$  \\
208 & $\f_i \fbar_i\to\stau_2 \stau_2^*$  \\
209 & $\f_i \fbar_i\to\stau_1 \stau_2^* +$ \\
210 & $\f_i \fbar_j\to \sell_L {\snu}_\ell^* +$\\
211 & $\f_i \fbar_j\to \stau_1\tilde{\nu}_\tau^* +$ \\
212 & $\f_i \fbar_j\to \stau_2\tilde{\nu}_\tau{}^* +$\\
213 & $\f_i \fbar_i\to \tilde{\nu_\ell} \tilde{\nu_\ell}^*$  \\
214 & $\f_i \fbar_i\to \tilde{\nu}_{\tau} \tilde{\nu}_{\tau}^*$ \\
216 & $\f_i \fbar_i \to \chio_1 \chio_1$  \\
217 & $\f_i \fbar_i \to \chio_2 \chio_2$  \\
218 & $\f_i \fbar_i \to \chio_3 \chio_3$  \\
219 & $\f_i \fbar_i \to \chio_4 \chio_4$  \\
220 & $\f_i \fbar_i \to \chio_1 \chio_2$  \\
221 & $\f_i \fbar_i \to \chio_1 \chio_3$  \\
222 & $\f_i \fbar_i \to \chio_1 \chio_4$  \\
223 & $\f_i \fbar_i \to \chio_2 \chio_3$  \\
224 & $\f_i \fbar_i \to \chio_2 \chio_4$  \\
225 & $\f_i \fbar_i \to \chio_3 \chio_4$  \\
226 & $\f_i \fbar_i \to \chip_1 \chim_1$  \\
227 & $\f_i \fbar_i \to \chip_2 \chim_2$  \\
228 & $\f_i \fbar_i \to \chip_1 \chim_2$  \\
229 & $\f_i \fbar_j \to \chio_1 \chip_1$  \\
\hline
\end{tabular}
\hfill
\begin{tabular}[t]{|rl|@{\protect\rule[-2mm]{0mm}{6mm}}}
\hline
No. & Subprocess \\ 
\hline
230 & $\f_i \fbar_j \to \chio_2 \chip_1$  \\
231 & $\f_i \fbar_j \to \chio_3 \chip_1$  \\
232 & $\f_i \fbar_j \to \chio_4 \chip_1$  \\
233 & $\f_i \fbar_j \to \chio_1 \chip_2$  \\
234 & $\f_i \fbar_j \to \chio_2 \chip_2$  \\
235 & $\f_i \fbar_j \to \chio_3 \chip_2$  \\
236 & $\f_i \fbar_j \to \chio_4 \chip_2$  \\
237 & $\f_i \fbar_i \to \glu \chio_1$    \\
238 & $\f_i \fbar_i \to \glu \chio_2$    \\
239 & $\f_i \fbar_i \to \glu \chio_3$    \\
240 & $\f_i \fbar_i \to \glu \chio_4$    \\
241 & $\f_i \fbar_j \to \glu \chip_1$    \\
242 & $\f_i \fbar_j \to \glu \chip_2$    \\
243 & $\f_i \fbar_i \to \glu \glu$\\
244 & $\g \g \to \glu \glu$\\
246 & $\f_i \g \to {\sq_i}{}_L \chio_1$\\
247 & $\f_i \g \to {\sq_i}{}_R \chio_1$\\
248 & $\f_i \g \to {\sq_i}{}_L \chio_2$\\
249 & $\f_i \g \to {\sq_i}{}_R \chio_2$\\
250 & $\f_i \g \to {\sq_i}{}_L \chio_3$\\
251 & $\f_i \g \to {\sq_i}{}_R \chio_3$\\
252 & $\f_i \g \to {\sq_i}{}_L \chio_4$\\
253 & $\f_i \g \to {\sq_i}{}_R \chio_4$\\
254 & $\f_i \g \to {\sq_j}{}_L \chip_1$\\
256 & $\f_i \g \to {\sq_j}{}_L \chip_2$\\
258 & $\f_i \g \to {\sq_i}{}_L \glu$ \\
259 & $\f_i \g \to {\sq_i}{}_R \glu$ \\
261 & $\f_i \fbar_i \to \tp_1 \tm_1$\\
262 & $\f_i \fbar_i \to \tp_2 \tm_2$\\
\hline
\end{tabular}
\hfill
\begin{tabular}[t]{|rl|@{\protect\rule[-2mm]{0mm}{6mm}}}
\hline
No. & Subprocess \\ 
\hline
263 & $\f_i \fbar_i \to \tp_1 \tm_2 +$\\
264 & $\g \g \to \tp_1 \tm_1$\\
265 & $\g \g \to \tp_2 \tm_2$\\
271 & $\f_i \f_j \to {\sq_i}{}_L {\sq_j}{}_L$\\
272 & $\f_i \f_j \to {\sq_i}{}_R {\sq_j}{}_R$\\
273 & $\f_i \f_j \to {\sq_i}{}_L {\sq_j}{}_R +$\\
274 & $\f_i \fbar_j \to {\sq_i}{}_L {\sqs_j}{}_L$\\
275 & $\f_i \fbar_j \to {\sq_i}{}_R {\sqs_j}{}_R$\\
276 & $\f_i \fbar_j \to {\sq_i}{}_L {\sqs_j}{}_R +$\\
277 & $\f_i \fbar_i \to {\sq_j}{}_L {\sqs_j}{}_L$\\
278 & $\f_i \fbar_i \to {\sq_j}{}_R {\sqs_j}{}_R$\\
279 & $\g \g \to {\sq_i}{}_L {\sqs_i}{}_L$\\
280 & $\g \g \to {\sq_i}{}_R {\sqs_i}{}_R$\\
281  & $\b \q_i \to \sbo_1 {\sq_i}{}_L$\\
282  & $\b \q_i \to \sbo_2 {\sq_i}{}_R$\\
283  & $\b \q_i \to \sbo_1 {\sq_i}{}_R + \sbo_2 {\sq_i}{}_L$\\
284  & $\b \qbar_i \to \sbo_1 {\sqs_i}{}_L$\\
285  & $\b \qbar_i \to \sbo_2 {\sqs_i}{}_R$\\
286  & $\b \qbar_i \to \sbo_1 {\sqs_i}{}_R + \sbo_2 {\sqs_i}{}_L$\\
287  & $\q_i \qbar_i \to \sbo_1 \sbs_1$\\
288  & $\q_i \qbar_i \to \sbo_2 \sbs_2$\\
289  & $\g \g \to \sbo_1 \sbs_1$\\
290  & $\g \g \to \sbo_2 \sbs_2$\\
291  & $\b \b \to \sbo_1 \sbo_1$\\
292  & $\b \b \to \sbo_2 \sbo_2$\\
293  & $\b \b \to \sbo_1 \sbo_2$\\
294  & $\b \g \to \sbo_1 \glu$\\
295  & $\b \g \to \sbo_2 \glu$\\
296  & $\b \bbar \to \sbo_1 \sbs_2 +$\\ 
\hline
\end{tabular}
\end{table}
 
\subsubsection{Process Classifications}
 
\textsc{Pythia} contains a rich selection of physics scenarios, with 
well above 200 different subprocesses, see Tables~\ref{table1}, 
\ref{table2} and \ref{table3}. The process number space has tended to 
become a bit busy, so processes are not always numbered logically.
Some processes are closely related variants of the same basic process 
(e.g. the production of a neutralino pair in processes 216--225),
others are alternative formulations (e.g. $\Z^0\Z^0 \to \hrm^0$
has to be convoluted with the flux of $\Z^0$'s around fermions, and thus 
is an approximation to $\f_i \f_j \to \f_i \f_j \hrm^0$; when the
latter was implemented the former was still kept). A process may
also have hidden further layers of processes (e.g. $\H^{\pm}$ can be 
produced in top decays, whichever way top is produced).

One classification of the subprocesses is according to the physics 
scenario. The following major groups may be distinguished:
\begin{Itemize}
\item Hard QCD processes, i.e. leading to jet production.
\item Soft QCD processes, such as diffractive and elastic scattering,
and minimum-bias events. Hidden in this class is also process 96,
which is used internally for the merging of soft and hard physics,
and for the generation of multiple interactions.
\item Heavy-flavour production, both open and hidden (i.e. as
bound states like the $\J/\psi$). Hadronization of open heavy flavour
will be discussed in section~\ref{sec:lowmassfrag}. Some new processes 
have been added for closed heavy flavour, but we remind that data here
are yet not fully understood, and have given rise to models extending
on the more conventional \textsc{Pythia} treatment \cite{clhflav}.
\item $\W / \Z$ production. A first-order process such as 
$\f_i \fbar_j \to \g \W^{\pm}$ is now quite accurately modeled
by the initial-state shower acting on $\f_i \fbar_j \to \W^{\pm}$,
see section~\ref{sec:inishow}, but the former can still be useful 
for a dedicated study of the high-$\pT$ tail.
\item Prompt-photon production.
\item Photon-induced processes, including Deep Inelastic Scattering.
A completely new machinery for $\gast\p$ and $\gast\gast$ physics has 
been constructed here, see section~\ref{sec:photon}.
\item Standard model Higgs production, where the Higgs is reasonably
light and narrow, and can therefore still be considered as a resonance.
\item Gauge boson scattering processes, such as 
$\W_{\mrm{L}} \W_{\mrm{L}} \to \W_{\mrm{L}} \W_{\mrm{L}}$ 
(L = longitudinal),  
when the standard model Higgs is so heavy and broad that resonant and
non-resonant contributions have to be considered together.
\item Non-standard Higgs particle production, within the framework
of a two-Higgs-doublet scenario with three neutral ($\hrm^0$, $\H^0$
and $\A^0$) and two charged ($\H^{\pm}$) Higgs states. Normally 
associated with \textsc{Susy} (see below), but does not have to be. 
The Higgs pair production processes were previously hidden in process 
141, but are now included explicitly. 
\item Production of new gauge bosons, such as a $\Z'$, $\W'$ and $\R$
(a horizontal boson, coupling between generations).
\item Technicolor production, as an alternative scenario to the 
standard picture of electroweak symmetry breaking by a fundamental Higgs. 
Processes 149, 191, 192 and 193 may be considered obsolete, since the 
other processes now include the decays to the allowed final states of 
the $\rho_{\mrm{tc}}^0/\omega_{\mrm{tc}}^0/\rho_{\mrm{tc}}^{\pm}$   
bosons, also including interferences with $\gamma/\Z^0/\W^{\pm}$.
The default scenario is based on \cite{techni}.
\item Compositeness is a possibility not only in the Higgs sector, 
but may also apply to fermions, e.g. giving $\d^*$ and $\u^*$ production.
At energies below the threshold for new particle production, contact
interactions may still modify the standard behaviour; this is implemented
not only for processes 165 and 166, but also for 11, 12 and 20.
\item Left--right symmetric models give rise to doubly charged Higgs
states, in fact one set belonging to the left and one to the right SU(2)
gauge group. Decays involve right-handed $\W$'s and neutrinos. 
The existing scenario is based on \cite{dchigg}.
\item Leptoquark ($\L_{\Q}$) production is encountered in some 
beyond-the-standard-model scenarios.
\item Supersymmetry (\textsc{Susy}) is probably the favourite scenario for
physics beyond the standard model. A rich set of processes are allowed,
even if one obeys $R$-parity conservation. The supersymmetric machinery 
and process selection is inherited from \textsc{SPythia}~\cite{SPYTHIA}, 
however with many improvements in the event generation chain. Relative 
to the \textsc{SPythia} process repertoire, the main new additions
is sbottom production, where a classification by mass eigenstates is 
necessary and many Feynman graphs are related to the possibility to have
incoming $\b$ quarks. Many different \textsc{Susy} scenarios have been 
proposed, and the program is flexible enough to allow input from several 
of these, in addition to the ones provided internally.
\end{Itemize}
Obviously the list is far from exhaustive; it is a major problem to 
keep up to date with all the new physics scenarios and signals that 
are proposed and have to be studied. One example of another physics
area that has attracted much attention recently is the possibility of
extra dimensions on `macroscopic' scales. Also, a general-purpose 
program can not be optimized for all kinds of processes. If a 
generator for some kind of partonic configurations is already available, 
outside of \textsc{Pythia}, there exists the possibility to feed this 
in for subsequent treatment of showers and hadronization. 

\subsubsection{Parton Distributions}
 
For cross section calculations, the hard partonic cross section has 
to be convoluted with the parton distributions of the incoming
beam particles. The current default is GRV~94L for protons \cite{GRV}
and SaS~1D for real and virtual photons \cite{SaSpdf}. Some further
parameterizations are available in \textsc{Pythia}, such as the recent
CTEQ~5 proton ones \cite{CTEQ}, and a much richer repertoire if the 
\textsc{Pdflib} library \cite{PDFLIB} is linked.
 
\subsubsection{Photon Physics}
\label{sec:photon}

Since before, a model for the interactions of real photons is 
available, i.e. for $\gamma\p$ and $\gamma\gamma$ events 
\cite{gagaSaS}. This has now been improved and extended also
to include virtual photons, i.e. $\gast\p$ and $\gast\gast$
events \cite{gagaFS}. It is especially geared towards the transition 
region of rather small photon virtualities $Q^2 \lessim 10$~GeV$^2$, 
where the physics picture is rather complex, while it may be overkill 
for large $Q^2$, where the picture again simplifies.

Photon interactions are complicated since the photon wave function
contains so many components, each with its own interactions. To
first approximation, it may be subdivided into a direct and a resolved
part. (In higher orders, the two parts can mix, so one has 
to provide sensible physical separations between the two.)
In the former the photon acts as a pointlike particle, 
while in the latter it fluctuates into hadronic states.
These fluctuations are of $\mathcal{O}(\alpha_{\mrm{em}})$, and so
correspond to a small fraction of the photon wave function, but this
is compensated by the bigger cross sections allowed in strong-interaction
processes. For real photons therefore the resolved processes dominate
the total cross section, while the pointlike ones take over for 
virtual photons. 
 
The fluctuations $\gamma \to \q\qbar \, (\to \gamma)$ can be characterized 
by the transverse momentum $\kT$ of the quarks, or alternatively by some
mass scale $m \simeq 2 \kT$, with a spectrum of fluctuations 
$\propto \d\kT^2/\kT^2$. The low-$\kT$ part cannot be calculated 
perturbatively, but is instead parameterized by experimentally determined  
couplings to the lowest-lying vector mesons, $V = \rho^0$, $\omega^0$, 
$\phi^0$ and $\J/\psi$, an ansatz called VMD for Vector Meson 
Dominance. Parton distributions are defined with a unit
momentum sum rule within a fluctuation \cite{SaSpdf}, giving rise
to total hadronic cross sections, jet activity, multiple interactions 
and beam remnants as in hadronic interactions. States at larger $\kT$
are called GVMD or Generalized VMD, and their contributions to the 
parton distribution of the photon are called anomalous. Given a dividing 
line $k_0 \simeq 0.5$~GeV to VMD states, the parton distributions are 
perturbatively calculable. The total cross section of a state is not, 
however, since this involves aspects of soft physics and eikonalization 
of jet rates. Therefore an ansatz is chosen where the total cross section 
scales like $k_V^2/\kT^2$, where the adjustable parameter 
$k_V \approx m_{\rho}/2$ for light quarks. The spectrum of states is taken 
to extend over a range $k_0 < \kT < k_1$, where $k_1$ is identified with 
the $p_{\perp\mathrm{min}}(s)$ defined in eq.~(\ref{eq:ptmin}) below. 
There is some arbitrariness in that choice, and for jet rate calculations 
also contributions to the parton distributions from above this region
are included. 

If the photon is virtual, it has a reduced probability to fluctuate into 
a vector meson state, and this state has a reduced interaction probability.
This can be modeled by a traditional dipole factor
$(m_V^2/(m_V^2 + Q^2))^2$ for a photon of virtuality $Q^2$, where 
$m_V \to 2 \kT$ for a GVMD state. Putting it all together, the cross
section of the GVMD sector then scales like
\begin{equation}
\int_{k_0^2}^{k_1^2} \frac{\d\kT^2}{\kT^2} \, \frac{k_V^2}{\kT^2} \,
\left( \frac{4\kT^2}{4\kT^2 + Q^2} \right)^2 ~.
\end{equation}
 
A real direct photon in a $\gamma\p$ collision can interact with the parton
content of the proton: $\gamma\q \to \q\g$ and $\gamma\g \to \q\qbar$.
The $\pT$ in this collision is taken to exceed $k_1$, in order to avoid
double-counting with the interactions of the GVMD states. For a virtual
photon the DIS (deeply inelastic scattering) process $\gast \q \to \q$
is also possible, but by gauge invariance its cross section must
vanish in the limit $Q^2 \to 0$. At large $Q^2$, the direct processes 
can instead be considered as the $\mathcal{O}(\alpha_{\mrm{s}})$ correction
to the lowest-order DIS process. The DIS $\gast\p$ cross section
is here proportional to the structure function $F_2 (x, Q^2)$ with the
Bjorken $x = Q^2/(Q^2 + W^2)$. Since normal parton distribution 
parameterizations are frozen below some $Q_0$ scale and therefore do not
obey the gauge invariance condition, an ad hoc factor 
$(Q^2/(Q^2 + m_{\rho}^2))^2$ is introduced for the conversion from 
the parameterized $F_2(x,Q^2) = \sum e_{\q}^2 \, q(x,Q^2)$ to a 
$\sigma_{\mrm{DIS}}^{\gast\p}$. In order to avoid double-counting
between DIS and direct events, we decide to introduce a requirement 
$\pT > \max(k_1, Q)$ on direct events. In the remaining DIS ones, thus
$\pT < Q$. The DIS rate should be reduced accordingly, by a Sudakov form 
factor giving the probability not to have an interaction above scale $Q$,
which can be approximated by 
$\exp(- \sigma_{\mrm{direct}}^{\gast\p} / \sigma_{\mrm{DIS}}^{\gast\p})$.

Note that the $Q^2$ dependence of the DIS and direct processes is 
implemented in the matrix element expressions. This is different from 
VMD/GVMD, where dipole factors are used to reduce the assumed flux of 
partons inside a virtual photon relative to those of a real one, but 
the matrix elements contain no parton virtuality dependence.

After some further minor corrections for double-counting, we arrive at a 
picture of hadronic $\gast\p$ events as being composed of four main 
components: VMD, GVMD, direct and DIS. Most of these in their turn have
a complicated internal structure, as we have seen. The $\gast\gast$ 
collision between two inequivalent photons contains 13 components: four 
when the VMD and GVMD states interact with each other (`double-resolved'),
eight with a direct or DIS photon interaction on a VMD or GVMD state on
either side (`single-resolved', including the traditional DIS), and one 
where two direct photons interact by the  process $\gast\gast \to \q\qbar$ 
(`direct', not to be confused with the direct process of $\gast\p$). 

Several further aspects can be added to the above machinery. The impact of
resolved longitudinal photons is unknown, except that it has to vanish in 
the limit $Q^2 \to 0$, and can be approximated by some $Q^2$-dependent 
enhancement of the normal transverse one. For a complete description of
$\e\p$ events or $\e^+\e^-$ two-photon ones, a convolution with the 
$x$- and $Q^2$-dependent flux of virtual photons inside an electron is 
also now provided.
 
\subsubsection{Supersymmetry}

\textsc{Pythia} simulates the Minimal Supersymmetric Standard Model (MSSM),
based on an effective Lagrangian of softly-broken \textsc{Susy}
with parameters defined at the weak scale, which is typically between
$m_{\Z}$ and 1 TeV.  The MSSM particle spectrum is minimal in the sense that
it includes only the partners
of all Standard Model particles (presently without massive neutrinos),
a two-Higgs doublet --- one Higgs ${\H}_{\u}$ coupling
only to up-type fermions and one ${\H}_{\d}$ coupling only to down-type 
fermions --- and partners, and the gravitino.
Once the parameters of the softly-broken \textsc{Susy} Lagrangian are
specified, the interactions are fixed, and the sparticle masses can
be calculated \cite{Haber}.
 
The masses of the scalar partners to fermions, sfermions,
depend on soft scalar masses, trilinear couplings, the Higgsino mass $\mu$,
and $\tan\beta$, the ratio of Higgs vacuum expectation values 
$\langle {\H}_{\u} \rangle / \langle {\H}_{\d} \rangle$.  The masses of the 
fermion partners to the gauge and Higgs bosons, the neutralinos and charginos, 
depend on soft gaugino masses, $\mu$, and $\tan\beta$.  Finally, the 
properties of the Higgs scalar sector is calculated from the input 
pseudoscalar Higgs boson mass $m_{\A}$, $\tan\beta$, $\mu$, trilinear 
couplings and the sparticle properties in an effective potential approach
\cite{Carena}. Of course, these calculations also depend on SM parameters 
($m_{\t}, m_{\Z}, \alpha_{\mrm{s}},$ etc.).  Any modifications to these
quantities from virtual MSSM effects are not taken into account.  
In principle, the sparticle masses also acquire loop corrections that 
depend on all MSSM masses.
 
$R$-parity conservation is assumed (at least on the time and distance
scale of a typical collider experiment), and
only lowest order, sparticle pair production processes are included.
Only those processes with $\e^+\e^-, \mu^+\mu^-$, or quark
and gluon initial states are simulated.  Likewise, only $R$-parity conserving
decays are allowed, so that one sparticle is stable, either the lightest
neutralino, the gravitino, or a sneutrino.
\textsc{Susy} decays of the top quark are included, but all other SM particle
decays are unaltered.
 
Various improvements to the simulation are being implemented in stages.
Some of these can have a significant impact on the collider phenomenology.
Among these are: the generalization to complex-valued soft 
\textsc{Susy}-breaking parameters in the neutralino and chargino sector; 
the same in the Higgs sector, which removes the possibility of CP-even or 
CP-odd labels; the calculation of neutralino and chargino decay rates which 
are accurate for large $\tan\beta$; and matrix element
weighting of particle distributions in three-body decays.
  
\subsubsection{Strong Dynamics in Electroweak Symmetry Breaking}

The simulation of strong dynamics associated with electroweak symmetry
breaking in \textsc{Pythia} is based on an effective Lagrangian for
the lightest resonances of a technicolor (TC)-like model.
In TC, the breaking of a chiral symmetry in a new, strongly interacting
gauge theory generates the Goldstone bosons necessary for
electroweak symmetry breaking.  Bound states of technifermions provide
a QCD-like spectrum of technipions ($\pi_{\mrm{tc}}$), technirhos 
($\rho_{\mrm{tc}}$), techniomegas ($\omega_{\mrm{tc}}$), etc. The mass 
hierarchies, however, are unlike QCD because of the behavior of the 
gauge couplings in realistic models of extended TC (ETC).  The difficulties 
of ETC in explaining the top quark mass while suppressing FCNC's is 
circumvented by the addition of topcolor interactions, which provide the 
bulk of $m_{\t}$.
 
In ETC models, hard mass contributions to technipion masses make
decays like $\rho_{\mrm{tc}} \to \pi_{\mrm{tc}}\pi_{\mrm{tc}}$ 
kinematically inaccessible.  Instead, decays like 
$\rho_{\mrm{tc}}^{\mrm{ew}} \to \pi_{\mrm{tc}}^{\mrm{ew}} {\W}_{\mrm{L}}$, 
for example, dominate, where $\mrm{ew}$ denotes constituent technifermions 
with only electroweak quantum numbers and ${\W}_{\mrm{L}}$
is a longitudinal $\W$ bosons.  As a result, the $\mrm{ew}$ technirho and
techniomega tend to have small total widths.
 
Effective couplings are derived in the valence technifermion approximation,
and the techniparticle decays can be calculated directly \cite{techni}.
Technirhos and techniomegas are produced through kinematic mixing 
with gauge bosons, leading to final states containing
Standard Model particles and/or pseudo-Goldstone bosons (technipions).
 
As an additional wrinkle, SU$_c$(3) non-singlet states are included
along with the coloron of topcolor assisted technicolor.  In this case,
colored technirhos (and the coloron) can have substantial total widths
and enhanced couplings to bottom and top quarks.
 
\subsection{QCD Radiation}

The matrix-element (ME) and parton-shower (PS) approaches to higher-order
QCD corrections both have their advantages and disadvantages. The former
offers a systematic expansion in orders of $\alpha_{\mrm{s}}$, and a powerful
machinery to handle multi-parton configurations on the Born level, 
but loop calculations are tough and lead to messy cancellations at
small resolution scales. Resummed matrix elements may circumvent
the latter problem for specific quantities, but then do not
provide exclusive accompanying events. Parton showers are based
on an improved leading-log (almost next-to-leading-log) approximation, 
and so cannot be accurate for well separated partons, but they offer a 
simple, process-independent machinery that gives a smooth blending of event 
classes (by Sudakov form factors) and a sensible match to hadronization.
It is therefore natural to try to combine these descriptions, so
that ME results are recovered for widely separated partons while the
PS sets the sub-jet structure. 

For final-state showers in $\Z^0 \to \q\qbar$, where $\q$ is assumed 
essentially massless, such solutions are the standard since long 
\cite{fsmatch}, e.g. by letting the shower slightly overpopulate the 
$\q\qbar\g$ phase space and then using a Monte Carlo veto technique to 
reduce down to the ME level. 

\subsubsection{Initial-State Showers}
\label{sec:inishow}

A similar technique is now available for the description of initial-state
radiation in the production of a single colour-singlet resonance, such
as $\gast/\Z^0/\W^{\pm}$ \cite{ismatch}. The basic idea is to map the 
kinematics between the PS and ME descriptions, and to find a correction 
factor that can be applied to hard emissions in the shower so as to bring
agreement with the matrix-element expression. The \textsc{Pythia} shower 
kinematics definitions are based on $Q^2$ as the spacelike virtuality of 
the parton produced in a branching and $z$ as the factor by which the 
$\hat{s}$ of the scattering subsystem is reduced by the branching. 
Some simple algebra then shows that the two $\q\qbar' \to \g\W^{\pm}$ 
emission rates disagree by a factor
\begin{equation}
R_{\q\qbar' \to \g\W}(\hat{s},\hat{t}) = 
\frac{(\mathrm{d}\hat{\sigma}/\mathrm{d}\hat{t})_{\mathrm{ME}} }%
     {(\mathrm{d}\hat{\sigma}/\mathrm{d}\hat{t})_{\mathrm{PS}} } = 
\frac{\hat{t}^2+\hat{u}^2+2 m_{\W}^2\hat{s}}{\hat{s}^2+m_{\W}^4} ~,
\label{RqqbargW} 
\end{equation}
which is always between $1/2$ and 1. 
The shower can therefore be improved in two ways, relative to the 
old description. Firstly, the maximum virtuality of emissions is 
raised from $Q^2_{\mathrm{max}} \approx m_{\W}^2$ to 
$Q^2_{\mathrm{max}} = s$, i.e. the shower is allowed to populate the 
full phase space. Secondly, the emission rate for the final (which 
normally also is the hardest) $\q \to \q\g$ emission on each side is 
corrected by the factor $R(\hat{s},\hat{t})$ above, so as to bring 
agreement with the matrix-element rate in the hard-emission region.
In the backwards evolution shower algorithm \cite{backwards}, this 
is the first branching considered.

The other possible ${\mathcal{O}}(\alpha_{\mrm{s}})$ graph is 
$\q\g \to \q'\W^{\pm}$, where the corresponding correction factor is
\begin{equation}
R_{\q\g \to \q'\W}(\hat{s},\hat{t}) =
\frac{(\mathrm{d}\hat{\sigma}/\mathrm{d}\hat{t})_{\mathrm{ME}} }%
     {(\mathrm{d}\hat{\sigma}/\mathrm{d}\hat{t})_{\mathrm{PS}} } = 
\frac{\hat{s}^2 + \hat{u}^2 + 2 m_{\W}^2 \hat{t}}{(\hat{s}-m_{\W}^2)^2 
+ m_{\W}^4} ~,
\end{equation}
which lies between 1 and 3. A probable reason for the lower shower 
rate here is that the shower does not explicitly simulate the $s$-channel 
graph $\q\g \to \q^* \to \q'\W$. The $\g \to \q\qbar$ branching 
therefore has to be preweighted by a factor of 3 in the shower, but 
otherwise the method works the same as above. Obviously, the shower 
will mix the two alternative branchings, and the correction factor 
for a final branching is based on the current type.

The reweighting procedure prompts some other changes in the shower. 
In particular, $\hat{u} < 0$ translates into a constraint on the phase
space of allowed branchings, not previously implemented. 

Our published comparisons with data on the $p_{\perp\W}$ spectrum 
show quite a good agreement with this improved simulation \cite{ismatch}. 
A worry was that an unexpectedly large primordial $k_{\perp}$, around 
4 GeV, was required to match the data in the low-$p_{\perp\W}$ region. 
However, at that time we had not realized that the data were not fully 
unsmeared. The required primordial $k_{\perp}$ therefore drops by about 
a factor of two \cite{primkTW}.

The method can also be used for initial-state photon emission, e.g. in 
the process $\e^+\e^- \to \gammaZ$. There the old default
$Q^2_{\mathrm{max}} = m_{\Z}^2$ allowed no emission at large $\pT$, 
$\pT \gtrsim m_{\Z}$ at LEP2. This is now corrected by the increased
$Q^2_{\mathrm{max}} = s$, and using the $R$ of eq.~(\ref{RqqbargW}) with
$m_{\W} \to m_{\Z}$.

The above method does not address the issue of next-to-leading 
order corrections to the total $\W$ cross section, which instead
can be studied with more sophisticated matching procedures
\cite{ismatchsteve}. Also extensions to other processes can be 
considered in the future.

There are also some other changes to the initial state radiation algorithm:
\begin{Itemize}
\item The cut on minimum gluon energy emitted in a branching is modified 
by an extra factor roughly corresponding to the $1/\gamma$ factor for the 
boost to the hard subprocess frame. Earlier, when a subsystem was strongly 
boosted, the minimum energy requirement became quite stringent on the 
low-energy incoming side, and could cut out much radiation. 
\item The angular-ordering requirement is now based on ordering $\pT/p$ rather 
than $\pT/p_L$, i.e. replacing $\tan\theta$ by $\sin\theta$. Earlier the 
starting value $(\tan\theta)_{\mrm{max}} = 10$ could actually be violated 
by some bona fide emissions for strongly boosted subsystems.
\item The $Q^2$ value of the backwards evolution of a heavy quark like $\crm$ 
in a proton beam is by force kept above $m_{\crm}^2$, so as to ensure that 
the branching $\g \to \crm \cbar$ is not `forgotten' by evolving $Q^2$ below 
$Q_0^2$. Thereby the possibility of having a $\crm$ in the beam remnant proper 
is eliminated \cite{smallmass}. The procedure is not forced for a photon beam, 
where charm occurs as part of the valence flavour content.
\item For incoming $\mu^{\pm}$ (or $\tau^{\pm}$) beams the kinematical 
variables are better selected to represent the differences in lepton mass,
and the lepton-inside-lepton parton distributions are properly defined. 
\end{Itemize}

\subsubsection{Final-State Showers}

The traditional final-state shower algorithm in \textsc{Pythia}
\cite{fsmatch} is based on an evolution in $Q^2 = m^2$, i.e. potential
branchings are considered in order of decreasing mass. A branching
$a \to bc$ is then characterized by $m_a^2$ and $z = E_b/E_a$.
For the process $\gammaZ \to \q\qbar$, the first gluon emission off  
both $\q$ and $\qbar$ are corrected to the first-order matrix elements
for $\gammaZ \to \q\qbar\g$. (The $\alphas$ and the Sudakov form
factor are omitted from the comparison, since the shower procedure here
attempts to include higher-order effects absent in the first-order
matrix elements.)

This matching is well-defined for massless quarks, and was originally
used unchanged for massive ones. A first attempt to include massive 
matrix elements did not compensate for mass effects in the shower
kinematics, and therefore came to exaggerate the suppression of 
radiation off heavy quarks \cite{QCDWG}. Now the shower has been modified 
to solve this issue, and also improved and extended to cover 
better a host of different reactions \cite{newfinshow}. 

The starting point is the calculation of processes
$a \to bc$ and $a \to bc\g$, where the ratio
\begin{equation}
W_{\mathrm{ME}}(x_1,x_2) =
\frac{1}{\sigma(a \to bc)} \, 
\frac{\d\sigma(a \to bc\g)}{\d x_1 \, \d x_2}
\end{equation} 
gives the process-dependent differential gluon-emission rate. 
Here the phase space variables are $x_1 = 2E_b/m_a$ and  
$x_2 = 2E_c/m_a$, expressed in the rest frame of parton $a$.
Using the standard model and the minimal supersymmetric extension
thereof as templates, a wide selection of colour and spin structures
have been addressed, exemplified by $\Z^0 \to \q\qbar$,
$\t \to \b\W^+$, $\H^0 \to \q\qbar$, $\t \to \b\H^+$,
$\Z^0 \to \sq\sqbar$, $\sq \to \sq'\W^+$, $\H^0 \to \sq\sqbar$,
$\sq \to \sq'\H^+$, $\chio \to \q\sqbar$, $\sq \to \q\chio$,
$\t \to \st\chio$, $\sg \to \q\sqbar$, $\sq \to \q\sg$, and
$\t \to \st\sg$. The mass ratios $r_1 = m_b / m_a$ and $r_2 = m_c/m_a$ 
have been kept as free parameters. When allowed, processes have
been calculated for an arbitrary mixture of ``parities'', i.e.
without or with a $\gamma_5$ factor, like in the vector/axial vector 
structure of $\gammaZ$. All the matrix elements are encoded in the new 
function \texttt{PYMAEL(NI,X1,X2,R1,R2,ALPHA)}, where \texttt{NI} 
distinguishes the matrix elements and \texttt{ALPHA} is related to the 
$\gamma_5$ admixture.

In order to match to the singularity structure of the massive matrix 
elements, the evolution variable $Q^2$ is changed from $m^2$ to 
$m^2 - m_{\mathrm{on-shell}}^2$, i.e. $1/Q^2$ is the propagator of a 
massive particle. Furthermore, the $z$ variable of a branching needs to 
be redefined, which is achieved by reducing the three-momenta
of the daughters in the rest frame of the mother. For the shower history 
$b \to b\g$ this gives a differential probability
\begin{equation}
W_{\mathrm{PS,1}}(x_1,x_2) 
= \frac{\alphas}{2\pi} \, C_F \, \frac{\d Q^2}{Q^2} \, 
\frac{2 \, \d z}{1-z} \, \frac{1}{\d x_1 \, \d x_2}
= \frac{\alphas}{2\pi} \, C_F \,
\frac{2}{x_3 \, (1 + r_2^2 - r_1^2 - x_2)}  ~,
\end{equation} 
where the numerator $1 + z^2$ of the splitting kernel for $\q \to \q \g$ 
has been replaced by a 2 in the shower algorithm. For a process with only 
one radiating parton in the final state, such as $\t \to \b\W^+$, the 
ratio $W_{\mathrm{ME}}/W_{\mathrm{PS,1}}$ gives the acceptance probability 
for an emission in the shower. The singularity structure exactly agrees 
between ME and PS, giving a well-behaved ratio always below unity. If both 
$b$ and $c$ can radiate, there is a second possible shower history that has 
to be considered. The matrix element is here split in two parts, one 
arbitrarily associated with $b \to b\g$ branchings and the other with 
$c \to c\g$ ones. A convenient choice is 
$W_{\mathrm{ME,1}} = W_{\mathrm{ME}} (1 + r_1^2 - r_2^2 - x_1)/x_3$ and 
$W_{\mathrm{ME,2}} = W_{\mathrm{ME}} (1 + r_2^2 - r_1^2 - x_2)/x_3$,
which again gives matching singularity structures in 
$W_{\mathrm{ME,}i}/W_{\mathrm{PS,}i}$ and thus a
well-behaved Monte Carlo procedure. 

Also subsequent emissions of gluons off the primary particles are 
corrected to $W_{\mathrm{ME}}$. To this 
end, a reduced-energy system is constructed, which retains the 
kinematics of the branching under consideration but omits the gluons 
already emitted, so that an effective three-body shower state can be 
mapped to an $(x_1, x_2, r_1, r_2)$ set of variables. For light quarks 
this procedure is almost equivalent with the original one of using the  
simple universal splitting kernels after the first branching. For
heavy quarks it offers an improved modelling of mass effects also in 
the collinear region.

Some further changes have been introduced, a few minor as default and
some more significant ones as non-default options \cite{newfinshow}. 
This includes the description of coherence effects and $\alphas$ 
arguments, in general and more specifically for secondary heavy flavour
production by gluon splittings.

Further issues remain to be addressed, e.g. radiation off particles
with non-negligible width. In general, however, the new shower should 
allow an improved description of gluon radiation in many different
processes.

\subsection{Beam Remnants and Multiple Interactions}

\subsubsection{Beam Remnants} 

In a hadron--hadron collision, the initial-state radiation algorithm
reconstructs one shower initiator in each beam, by backwards
evolution from the hard scattering. This initiator only
takes some fraction of the total beam energy, leaving behind a beam
remnant that takes the rest. Since the initiator is coloured, so is 
the remnant. It is therefore colour-connected to the hard interaction, 
and forms part of the same fragmenting system. Often the remnant can 
be complicated, e.g. a $\g$ initiator would leave behind a $\u \u \d$ 
proton-remnant system in a colour octet state, which can conveniently 
be subdivided into a colour triplet quark and a colour antitriplet 
diquark, each of which are colour-connected to the hard interaction. 
The energy sharing between these two remnant objects, and their 
relative transverse momentum, introduces additional nonperturbative
degrees of freedom. Some of the default values have recently
been updated \cite{smallmass}.
 
One would expect an $\ep$ event to have only one beam
remnant, and an $\ee$ event none. This is not always correct, e.g.
a $\gamma \gamma \to \q \qbar$ interaction in an $\ee$ event would
leave behind the $\e^+$ and $\e^-$ as beam remnants. The photons
may in their turn leave behind remnants.
 
It is customary to assign a primordial transverse momentum to the 
shower initiator, to take into account the motion of quarks inside 
the original hadron, basically as required by the uncertainty principle.
A number of the order of 
$\langle \kT \rangle \approx m_{\p}/3 \approx 300$~MeV
could therefore be expected. However, in hadronic collisions much 
higher numbers than that are often required to describe data, 
typically of the order of 1 GeV \cite{primkT,primkTW} if
a Gaussian parameterization is used. (This number is now the default.) 
Thus, an interpretation as a purely nonperturbative motion 
inside a hadron is difficult to maintain. 

Instead a likely culprit is the initial-state shower algorithm. This 
is set up to cover the region of hard emissions, but may miss out on 
some of the softer activity, which inherently borders on 
nonperturbative physics. By default, the shower does not evolve down 
to scales below $Q_0 = 1$~GeV. Any shortfall in shower activity around 
or below this cutoff then has to be compensated by the primordial 
$\kT$ source, which thereby largely loses its original meaning.

\subsubsection{Multiple Interactions}

Multiple parton--parton interactions is the concept that, based on
the composite nature of hadrons, several parton pairs may interact
in a typical hadron--hadron collision \cite{multint}. Over the years, 
evidence for this mechanism has accumulated, such as the recent direct
observation by CDF \cite{cdfmultint}. The occurrences with two 
parton pairs at reasonably large $\pT$ just form the top of
the iceberg, however. In the \textsc{Pythia} model, most 
interactions are at lower $\pT$, where they are not visible as 
separate jets but only contribute to the underlying event structure. 
As such, they are at the origin of a number of key features, like the 
broad multiplicity distributions, the significant forward--backward
multiplicity correlations, and the pedestal effect under jets.

Since the perturbative jet cross section is divergent for 
$\pT \to 0$, it is necessary to regularize it, e.g. by a
cut-off at some $p_{\perp\mathrm{min}}$ scale. That such a 
regularization should occur is clear from the fact that the incoming
hadrons are colour singlets --- unlike the coloured partons assumed in 
the divergent perturbative calculations --- and that therefore the 
colour charges should screen each other in the $\pT \to 0$ limit.
Also other damping mechanisms are possible \cite{dampmialt}.
Fits to data typically give $p_{\perp\mathrm{min}} \approx 2$ GeV,
which then should be interpreted as the inverse of some colour
screening length in the hadron.   

One key question is the energy-dependence of $p_{\perp\mathrm{min}}$; 
this may be relevant e.g. for comparisons of jet rates at different 
Tevatron energies, and even more for any extrapolation to LHC energies. 
The problem actually is more pressing now than at the time of the 
original study \cite{multint}, since nowadays parton distributions are 
known to be rising more steeply at small $x$ than the flat $xf(x)$ 
behaviour normally assumed for small $Q^2$ before HERA. This 
translates into a more dramatic energy dependence of the 
multiple-interactions rate for a fixed $p_{\perp\mathrm{min}}$. 

The larger number of partons also should increase the amount of
screening, however, as confirmed by toy simulations \cite{dampmi}.
As a simple first approximation, $p_{\perp\mathrm{min}}$ is assumed
to increase in the same way as the total cross section, i.e. with some 
power $\epsilon \approx 0.08$ \cite{reggeondl} that, via reggeon 
phenomenology, should relate to the behaviour of parton distributions 
at small $x$ and $Q^2$. Thus the new default in PYTHIA is
\begin{equation}
p_{\perp\mathrm{min}}(s) = (1.9~{\mathrm{GeV}}) \left(
\frac{s}{1~\mathrm{TeV}^2} \right)^{0.08} ~.
\label{eq:ptmin}
\end{equation}
 
\subsection{Fragmentation and Decays}
 
QCD perturbation theory, formulated in terms of quarks and
gluons, is valid at short distances. At long distances, QCD
becomes strongly interacting and perturbation theory breaks
down. In this confinement r\'egime, the coloured partons are
transformed into colourless hadrons, a process called either
hadronization or fragmentation. 
 
The fragmentation process has yet to be understood from first
principles, starting from the QCD Lagrangian. This has left the
way clear for the development of a number of different
phenomenological models. \textsc{Pythia} is intimately connected 
with string fragmentation, in the form of the time-honoured 
`Lund model' \cite{lundfrag}. This is the default for all 
applications. Improvements have been made in some areas, however.

\subsubsection{Low-Mass Strings} 
\label{sec:lowmassfrag}

A hadronic event is conventionally subdivided into sets of partons
that form separate colour singlets. These sets are represented by strings,
that e.g. stretch from a quark end via a number of intermediate gluons
to an antiquark end. Three string mass regions may be distinguished for 
the hadronization.
\begin{Enumerate}
\item {\em Normal string fragmentation}. In the ideal situation, each 
string has a large invariant mass. Then the standard iterative 
fragmentation scheme \cite{lundfrag,multstring} works well. In practice, 
this approach can be used for all strings above some cut-off mass of a 
few GeV. 
\item {\em Cluster decay}.
If a string is produced with a \mbox{small} invariant mass, maybe only 
two-body final states are kinematically accessible. The traditional 
iterative Lund scheme is then not applicable. We call such a low-mass 
string a cluster, and consider it separately from above. In recent 
program versions, the modeling has now been improved to give a smooth 
match on to the standard string scheme in the high-cluster-mass limit
\cite{smallmass}.
\item {\em Cluster collapse}.
This is the extreme case of the above situation, where the string 
mass is so small that the cluster cannot decay into two hadrons.
It is then assumed to collapse directly into a single hadron, which
inherits the flavour content of the string endpoints. The original 
continuum of string/cluster masses is replaced by a discrete set
of hadron masses. Energy and momentum then cannot be conserved
inside the cluster, but must be exchanged with the rest of the event.
This description has also been improved \cite{smallmass}.  
\end{Enumerate}

String systems below a threshold mass are handled by the cluster 
machinery. In it, an attempt is first made to produce two hadrons, 
by having the string break in the middle by the production of a new 
$\q\qbar$ pair, with flavours and hadron spins selected according to 
the normal string rules. If the sum of the hadron masses 
is larger than the cluster mass, repeated attempts can be made to find
allowed hadrons; the default is two tries. If an allowed set is found,
the angular distribution of the decay products in the cluster rest
framed is picked isotropically near the threshold, but then gradually
more elongated along the string direction, to provide a smooth match
to the string description at larger masses. This also includes a 
forward--backward asymmetry, so that each hadron is preferentially in 
the same hemisphere as the respective original quark it inherits.

If the attempts to find two hadrons fail, one single hadron is formed 
from the given flavour content. The basic strategy thereafter is to 
exchange some minimal amount of energy and momentum between the
collapsing cluster and other string pieces in the neighbourhood.
The momentum transfer can be in either direction, depending on 
whether the hadron is lighter or heavier than the cluster it comes 
from. When lighter, the excess momentum is split off and put as an
extra `gluon' on the nearest string piece, where `nearest' is 
defined by a space--time history-based distance measure. When the 
hadron is heavier, momentum is instead borrowed from the endpoints 
of the nearest string piece.

The free parameters of the model can be tuned to data, especially
to the significant asymmetries observed between the production of
$\D$ and $\Dbar$ mesons in $\pi^- \p$ collisions, with hadrons
that share some of the $\pi^-$ flavour content very much favoured at 
large $x_F$ in the $\pi^-$ fragmentation region \cite{casym}. 
These spectra and asymmetries are closely related to the cluster 
collapse mechanism, and also to other effects of the colour topology
of the event (`beam drag') \cite{smallmass}. Also other parameters
enter the description, however, such as the effective charm mass and 
the beam remnant structure.

\subsubsection{Baryon Production}

A new advanced scheme has been introduced for baryon production
with the popcorn mechanism \cite{newbaryon}, plus some minor changes 
to the older popcorn scheme \cite{oldbaryon}. These new features 
currently only appear as options, with the default unchanged,
and can be separated into three parts.

Firstly, an improved implementation of SU(6) weights for baryon 
production. This should not be regarded as a new model, rather a more 
correct implementation of the old. However, in order to enable the user 
to see the effects of the SU(6) weighting separately, both procedures 
are available as different options. The main change is that, if a
step $\q \to \B + \qbar\qbar'$ is SU(6)-rejected, the new try may now
instead give a $\q \to \M + \q'$ step (where $\B$ stands for baryon, 
$\M$ for meson). The old procedure leads to a 
slightly faster algorithm and a better interpretation of the input 
parameter for the diquark-to-quark production rate. However, the 
probability that a quark will produce a baryon and a antidiquark is 
then flavour independent, which is not in agreement with the model.
Further, for $\q\q \to \M + \q\q'$, SU(6) symmetry is included in the 
weights for $\q\q'$, while $\q\q$ is kept with unit probability.
The procedures for $\q\q \to \B + \qbar'$ and a final joining 
$\q\q + \q \to \B$ are unchanged.

Secondly, a suppression of diquark vertices occuring at small proper 
times. This is based on a study of the production dynamics of the 
three quarks that form a baryon. The main experimental consequence is 
a suppression of the baryon production rate at large momentum fraction.
This in particular implies a smaller rate of first-rank light baryon 
production, while charm and bottom baryons are less affected
(since the production proper time is larger for a heavy hadron
than a light one of the same momentum). It thereby 
substitutes and explains the older brute-force possibility to 
suppress the production of first-rank baryons. 

Thirdly, a completely new flavour algorithm for baryons and popcorn
mesons, also using the small-proper-time suppression above. While the old 
popcorn alternative allowed at most one meson to be produced in between 
the baryon and the antibaryon, the new model allows an arbitrary number. 
The new flavour model makes explicit use of the popcorn suppression 
factor $\exp(-2 m_{\perp} M_{\perp}/\kappa)$, where $m_{\perp}$ is the 
transverse mass of the quark creating the colour fluctuation, 
$M_{\perp}$ is the total invariant transverse mass of the popcorn 
meson system, and $\kappa$ is the string tension constant. Thus two 
parameters, representing the mean $2 m_{\perp}/\kappa$ for light quarks 
and $\s$-quarks, respectively, govern both diquark and popcorn meson 
production. A corresponding parameter is introduced for the fragmentation 
of strings that contain diquarks already from the beginning, i.e. baryon 
remnants. The new procedure therefore requires far fewer parameters than
the old one, and still provides a comparable quality in the description 
of the various baryon production rates. This was investigated in detail 
in \cite{JGTbaryon}. (The concluding worry of an ``improper treatment''
was caused by an unfortunate misunderstanding and can be disregarded.)
Other features, such as baryon correlations, are also modified.

Several new routines have been added, and the diquark code has
been extended with information about the curtain quark flavour, i.e.
the $\q\qbar$ pair that is shared between the baryon and antibaryon,
but this is not visible externally. Some parameters are no longer
used, while others have to be given modified values, as described in the
long writeup. 

\subsubsection{Interconnection Effects}

The widths of the $\W$, $\Z$ and $\t$ are all of the order of 
2 GeV. A standard model Higgs with a mass above 200 GeV, as well 
as many supersymmetric and other beyond the standard model particles
would also have widths in the multi-GeV range. Not far from
threshold, the typical decay times 
$\tau = 1/\Gamma  \approx 0.1 \, {\mathrm{fm}} \ll  
\tau_{\mathrm{had}} \approx 1 \, \mathrm{fm}$.
Thus hadronic decay systems overlap, between a resonance and the
underlying event, or between pairs of resonances, so that the final 
state may not contain independent resonance decays.

So far, studies have mainly been performed in the context of
$\W$ pair production at LEP2. Pragmatically, one may here distinguish 
three main eras for such interconnection:
\begin{Enumerate}
\item Perturbative: this is suppressed for gluon energies 
$\omega > \Gamma$ by propagator/timescale effects; thus only
soft gluons may contribute appreciably.
\item Nonperturbative in the hadroformation process:
normally modeled by a colour rearrangement between the partons 
produced in the two resonance decays and in the subsequent parton
showers.
\item Nonperturbative in the purely hadronic phase: best exemplified 
by Bose--Einstein effects.
\end{Enumerate}
The above topics are deeply related to the unsolved problems of 
strong interactions: confinement dynamics, $1/N^2_{\mathrm{C}}$ 
effects, quantum mechanical interferences, etc. Thus they offer 
an opportunity to study the dynamics of unstable particles,
and new ways to probe confinement dynamics in space and 
time \cite{GPZrec,KSrec}, {\em but} they also risk 
to limit or even spoil precision measurements.

The reconnection scenarios outlined in \cite{KSrec} are now
available, plus also an option along the lines suggested in
\cite{GHrec}. Currently they can only be invoked in process 25,
$\e^+\e^- \to \W^+\W^- \to \q_1\qbar_2\q_3\qbar_4$, which is the most
interesting one for the foreseeable future. (Process 22,
$\e^+\e^- \to \gammaZ \; \gammaZ \to 
\q_1\qbar_2\q_3\qbar_4$ can also be used, but the travel distance
is calculated based only on the $\Z^0$ propagator part.)
If normally the
event is considered as consisting of two separate colour singlets,
$\q_1\qbar_2$ from the $\W^+$ and $\q_3\qbar_4$ from the $\W^-$,
a colour rearrangement can give two new colour singlets 
$\q_1\qbar_4$ and $\q_3\qbar_2$. It therefore leads to a different
hadronic final state, although differences usually turn out to
be subtle and difficult to isolate \cite{recsearch}. When also
gluon emission is considered, the number of potential reconnection
topologies increases. Apart from the overall rate of reconnection,
the scenarios in \textsc{Pythia}  differ in the relative probability 
assigned to each of these topologies, based on their properties
in momentum space and/or space--time.  For instance, scenario I
is based on an analogy with type I superconductors, with the
colour field represented by extended flux tubes. By contrast,
scenario II assumes that narrow vortex lines carry all the 
topological information, like in type II superconductors, even if
the full energy is stored over a wider region. 

Bose--Einstein effects are simulated in a simplified manner,
by introducing small momentum shifts in identical final-state
mesons (primarily $\pi^{\pm}$ and $\pi^0$) so as to bring them 
closer to each other \cite{LSBE}. The shifts can be chosen  
to reproduce a desired BE enhancement shape for small relative 
momentum $Q = \sqrt{m_{ij}^2 - 4 m_i^2}$ between identical bosons
$i$ and $j$. Typically the shape is chosen as a Gaussian,
$f_2(Q) = 1 + \lambda \exp(-Q^2 R^2)$, with $\lambda$ and
$R$ two free parameters. The input is only exactly reproduced
in the limit of an isotropic and low-density initial particle 
distribution; since these conditions are not 
completely fulfilled in reality, there are distortions \cite{MSLBE}, 
for better or worse. (The nontrivial three-particle correlations 
in data are described qualitatively, although not quantitatively.) 

A major shortcoming of the algorithm is that energy is not 
automatically conserved, even though three-momentum is. In the 
original algorithm, this was solved by a uniform rescaling of all 
three-momenta, with undesirable side effects e.g. when studying BE 
effects in $\W^+\W^-$ hadronic final states. In the current version,
several new options have been added that, based on different principles,
instead shifts pairs apart. The default one, BE$_{32}$, operates on 
identical particles, introducing an extra factor
\begin{equation}
1 + \alpha \lambda \exp(-Q^2 R^2 /9) \left\{ 1 - \exp(-Q^2 R^2 /4)
\right\} 
\end{equation}
to $f_2(Q)$. Here $\alpha$ is a negative number adjusted event by event 
for overall energy conservation, with $\langle \alpha \rangle 
\approx -0.25$. This scenario can be viewed as a simplified version of 
a dampened oscillating correlation function, where only the first peak 
and dip has been retained. Further new options have also been introduced 
specifically geared towards studies of $\W^+\W^-$ hadronic events, e.g. 
to include the effects of the separated $\W^+$ and $\W^-$ decay vertices.
 
\subsubsection{Decays}
  
Two separate decay treatments exist in \textsc{Pythia}. One is making 
use of a set of tables where branching ratios and decay modes are
stored, and is used e.g. for hadronic decays, where branching
ratios normally cannot be calculated from first principles.

The other treatment is used for a set of fundamental resonances 
in or beyond the standard model, such as $\t$, $\Z^0$, $\W^{\pm}$, 
$\hrm^0$, supersymmetric particles, and many more. Characteristic
here is that these resonances have perturbatively calculable 
widths to each of their decay channels. The decay products
are typically quarks, leptons, or other resonances. In decays to 
quarks, parton showers are automatically added to give a more 
realistic multijet structure, and one may also allow photon 
emission off leptons. If the decay products in turn are resonances, 
further decays are necessary. Often spin information is available 
in resonance decay matrix elements, leading to nonisotropic decays.
This part has been improved in several processes, but is still missing 
in many others.

The routine used to calculate the partial and total width of resonances 
(now expressed in GeV throughout), has been expanded for all the new 
particles and decay modes introduced. Some alternative calculation 
schemes have also been adopted, e.g. based on a simple rescaling of 
the on-shell widths rather than a complete recalculation (which may 
at times not be feasible) based on the current mass.

The width to be used in the denominator of a resonance propagator is only
well-defined near the peak. Well away from the peak, an unfortunate
choice may lead to a loss of cancellation between resonant and
nonresonant diagrams. A special problem exists for a massive standard 
model Higgs, where the width $\Gamma_{\hrm} \propto m_{\hrm}^3$ is so 
large that the choice of $\hat{s}$ dependence of the width significantly
influences the resonance peak shape. Following \cite{hwidth},
the default now is $\Gamma_{\hrm} \propto m_{\hrm}^2\sqrt{\hat{s}}$.

\section{Program News}

Essentially all of the basic philosophy and framework remain from the 
previous \textsc{Pythia} and \textsc{Jetset} versions, so no user 
familiar with these should feel at loss with \textsc{Pythia}~6.1. 
Most of the changes and additions instead are under the surface, and 
are only visible as new options added to the existing repertoire. 
However, some changes are fairly obvious, and other less obvious 
ones still of general interest. These will be covered in this section, 
in fairly general terms. Again we refer to the \textsc{Pythia} web page 
for a detailed documentation.
 
\subsection{Coding conventions}

As before, the \textsc{Fortran}~77 standard is adhered to. A very few
minor extensions may be used in isolated places, like the 7-character 
names of the \textsc{Pdflib} routines \cite{PDFLIB}, but are not known 
to cause problems on any compiler in use.

An obvious consequence of the \textsc{Pythia}/\textsc{Jetset} code 
merging is that the old \textsc{Jetset} routines and commonblocks have 
been renamed to begin with \texttt{PY} (instead of \texttt{LU} or 
\texttt{UL}), just like the \textsc{Pythia} ones. In most cases, 
the rest of the name is unchanged, but 
there are a few exceptions, mainly \texttt{RLU}$\to$\texttt{PYR},
\texttt{KLU}$\to$\texttt{PYK}, \texttt{PLU}$\to$\texttt{PYP} and
\texttt{LUXTOT}$\to$\texttt{PYXTEE}. Three integer functions now
begin with \texttt{PY}, namely \texttt{PYK}, \texttt{PYCHGE} and
\texttt{PYCOMP}, and therefore have to be declared extra. The 
\texttt{LUDATA} block data has been merged into \texttt{PYDATA},
and the test routine \texttt{LUTEST} into \texttt{PYTEST}.
For rotations and boosts, the \texttt{PYROBO} routine now requires
the range of affected entries to be given, like the old
\texttt{LUDBRB} but unlike \texttt{LUROBO} (but 0,0 as range arguments 
gives back the old \texttt{LUROBO} behaviour).

All real variables are now in \texttt{DOUBLE PRECISION}, which is
assumed to mean 64 bits, and also real constants have been promoted
to the higher precision. This is required to ensure proper functioning 
at currently studied energies, such as the LHC and beyond. 
To take into account this, all routines begin with the declarations
\begin{verbatim} 
C...Double precision and integer declarations.
      IMPLICIT DOUBLE PRECISION(A-H, O-Z)
      IMPLICIT INTEGER(I-N)
      INTEGER PYK,PYCHGE,PYCOMP
\end{verbatim} 
and users should do the same in their main programs.

On a machine where \texttt{DOUBLE PRECISION} would give 128 bits,
it may make sense to use compiler options to revert to 64 bits,
since the program is anyway not constructed to make use of 128
bit precision. 

The random number generator is the same as in previous versions
\cite{marsaglia}, but has now been expanded to operate with a
48 bit mantissa for the real numbers.

\textsc{Fortran}~77 makes no provision for 
double-precision complex numbers, but since \texttt{COMPLEX} is 
used only sparingly, no problems should be expected from this
omission. For the technicolor processes, some variables are declared
\texttt{COMPLEX*16} in the \texttt{PYSIGH} routine. Should the 
compiler not accept this, that one declaration can be changed to
\texttt{COMPLEX} with some drop in precision for the affected 
processes.

\begin{table}[t]
\caption{The current form of the main commonblock declarations.
\protect\label{table4} } 
\begin{verbatim}
COMMON/PYJETS/N,NPAD,K(4000,5),P(4000,5),V(4000,5)
COMMON/PYDAT1/MSTU(200),PARU(200),MSTJ(200),PARJ(200)
COMMON/PYDAT2/KCHG(500,4),PMAS(500,4),PARF(2000),VCKM(4,4)
COMMON/PYDAT3/MDCY(500,3),MDME(4000,2),BRAT(4000),KFDP(4000,5)
COMMON/PYDAT4/CHAF(500,2)
CHARACTER CHAF*16
COMMON/PYDATR/MRPY(6),RRPY(100)
COMMON/PYSUBS/MSEL,MSELPD,MSUB(500),KFIN(2,-40:40),CKIN(200)
COMMON/PYPARS/MSTP(200),PARP(200),MSTI(200),PARI(200)
COMMON/PYINT1/MINT(400),VINT(400)
COMMON/PYINT2/ISET(500),KFPR(500,2),COEF(500,20),ICOL(40,4,2)
COMMON/PYINT3/XSFX(2,-40:40),ISIG(1000,3),SIGH(1000)
COMMON/PYINT4/MWID(500),WIDS(500,5)
COMMON/PYINT5/NGENPD,NGEN(0:500,3),XSEC(0:500,3)
COMMON/PYINT6/PROC(0:500)
CHARACTER PROC*28
COMMON/PYMSSM/IMSS(0:99),RMSS(0:99)
COMMON/PYUPPR/NUP,KUP(20,7),NFUP,IFUP(10,2),PUP(20,5),Q2UP(0:10)
COMMON/PYBINS/IHIST(4),INDX(1000),BIN(20000)
\end{verbatim}
\end{table} 

Several compilers report problems when an odd number of integers
precede a double-precision variable in a commonblock. Therefore
an extra integer has  been introduced as padding in a few instances
(\texttt{NPAD}, \texttt{MSELPD} and \texttt{NGENPD} in 
Table~\ref{table4}). 

In order to cater for the increased offering of 
subprocesses, some arrays in commonblocks have been expanded. A few, 
such as \texttt{PYINT4}, have also been reorganized to represent 
improvements in the physics modeling. Most commonblocks and commonblock
variables are easily recognizable from previous program versions,
however. The current complement is given in Table~\ref{table4},
omitting some of the less interesting ones.

Since \textsc{Fortran}~77 provides no date-and-time routine, 
\texttt{PYTIME} allows a system-specific routine to be interfaced,
with some commented-out examples given in the code.
This routine is only used for cosmetic improvements of the output,
however, so can be left at the default with time 0 given.

For a program written to run \textsc{Pythia}~5 and \textsc{Jetset}~7,
most of the conversion required for \textsc{Pythia}~6 is fairly 
straightforward, and can be automatized. Both a simple \textsc{Fortran}
routine and a more sophisticated \textsc{Perl} \cite{convperl} script 
exists to this end. Some manual checks and interventions may still 
be required.

\subsection{Particle codes and data}

\begin{table}[t]
\caption{New or modified particle codes or names.
\protect\label{table5} } 
\begin{tabular}[t]{|rl|@{\protect\rule[-2mm]{0mm}{6mm}}}
\hline
\multicolumn{2}{|l|@{\protect\rule[-2mm]{0mm}{7mm}}}{Renamed:} \\
  7 & $\b'$ \\
  8 & $\t'$ \\
 17 & $\tau'$ \\
 18 & $\nu_{\tau}'$ \\
 25 & $\hrm^0$ \\
 35 & $\H^0$ \\
\hline
\multicolumn{2}{|l|@{\protect\rule[-2mm]{0mm}{7mm}}}{Moved:} \\
  100443 & $\psi'$ \\
  100553 & $\Upsilon'$ \\
 4000001 & $\d^*$ \\
 4000002 & $\u^*$ \\
 4000011 & $\e^*$ \\
 4000012 & $\nu_{\e}^*$ \\
\hline
\end{tabular}
\hfill
\begin{tabular}[t]{|rl|@{\protect\rule[-2mm]{0mm}{6mm}}}
\hline
\multicolumn{2}{|l|@{\protect\rule[-2mm]{0mm}{7mm}}}{Technicolor:} \\
 51 & $\pi_{\mathrm{tc}}^0$ \\
 52 & $\pi_{\mathrm{tc}}^+$ \\
 53 & ${\pi'_{\mathrm{tc}}}^0$ \\
 54 & $\rho_{\mathrm{tc}}^0$ \\
 55 & $\rho_{\mathrm{tc}}^+$ \\
 56 & $\omega_{\mathrm{tc}}^0$ \\
\hline
\multicolumn{2}{|l|@{\protect\rule[-2mm]{0mm}{7mm}}}{LR-symmetric:} \\
 61 & $\H_L^{++}$ \\
 62 & $\H_R^{++}$ \\
 63 & $\W_R^+$ \\
 64 & $\nu_{R\e}$ \\
 65 & $\nu_{R\mu}$  \\
 66 & $\nu_{R\tau}$  \\
\hline
\end{tabular}
\hfill
\begin{tabular}[t]{|rl|@{\protect\rule[-2mm]{0mm}{6mm}}}
\hline
\multicolumn{2}{|l|@{\protect\rule[-2mm]{0mm}{7mm}}}{\textsc{Susy}:} \\
 1000001 & $\sd_L$ \\
 1000002 & $\su_L$ \\
 1000003 & $\sst_L$ \\
 1000004 & $\sch_L$ \\
 1000005 & $\sbo_1$ \\
 1000006 & $\st_1$ \\
 1000011 & $\se_L$ \\
 1000012 & $\tilde{\nu}_{{\e}L}$ \\
 1000013 & $\smu_L$ \\
 1000014 & $\tilde{\nu}_{{\mu}L}$ \\
 1000015 & $\stau_1$ \\
 1000016 & $\tilde{\nu}_{{\tau}L}$ \\
 1000021 & $\glu$ \\
 1000022 & $\chio^0_1$ \\
 1000023 & $\chio^0_2$ \\
 1000024 & $\chio^+_1$ \\
\hline
\end{tabular}
\hfill
\begin{tabular}[t]{|rl|@{\protect\rule[-2mm]{0mm}{6mm}}}
\hline
\multicolumn{2}{|l|@{\protect\rule[-2mm]{0mm}{7mm}}}{\textsc{Susy}:} \\
 2000001 & $\sd_R$ \\
 2000002 & $\su_R$ \\
 2000003 & $\sst_R$ \\
 2000004 & $\sch_R$ \\
 2000005 & $\sbo_2$ \\
 2000006 & $\st_2$ \\
 2000011 & $\se_R$ \\
 2000012 & $\tilde{\nu}_{{\e}R}$ \\
 2000013 & $\smu_R$ \\
 2000014 & $\tilde{\nu}_{{\mu}R}$ \\
 2000015 & $\stau_2$ \\
 2000016 & $\tilde{\nu}_{{\tau}R}$ \\
 1000025 & $\chio^0_3$ \\
 1000035 & $\chio^0_4$ \\
 1000037 & $\chio^+_2$ \\
 1000039 & $\grav$ \\
\hline
\end{tabular}
\end{table}

A number of new particle codes \texttt{KF} have been introduced, or 
modified, see Table~\ref{table5}. Mostly this is based on the PDG-agreed 
conventions \cite{PDG,LEP2QCD}, but some not yet standardized codes 
appear in the `empty' range 41--80. Furthermore, the fourth generation 
fermions and neutral scalar Higgs states have been renamed.
The two fermion spartners are labelled left and right, except in
the third generation, where an expected larger mixing makes
the two mass eigenstates a better choice of classification. 

The top hadrons are gone. It is now known that top is too short-lived 
to form hadronic bound states, so a reasonable description is instead 
to have the top quarks decay before hadronization is considered. The 
same is now assumed about a hypothetical fourth generation. Should the
need ever arise in the future to consider a new long-lived coloured
object, an effective description of a hadron as a small string with 
an ordinary colour-matching flavour at the other end should be sufficient.
One such example would be leptoquark-hadrons \cite{lqhad}.

Bottom hadrons are now defined individually, e.g. the previous common
decay scheme is gone in favour of individual branching ratios for each
hadron. On the other hand, given the sketchy knowledge of many branching 
ratios, the default description is still fairly standardized. 

Decay data is mainly based on the 1996 PDG edition \cite{PDGold}, but 
with many `educated guesses' to fill in missing information.

Since running fermion masses are used in an increasing number of processes,
e.g. for Higgs couplings, a function \texttt{PYMRUN(KF,Q2)} has been 
introduced to give the mass as a function of $\Q^2$ scale.

The compressed codes, \texttt{KC = PYCOMP(KF)}, are completely 
changed. We remind that \texttt{KF} can range up to seven-digit codes,
plus a sign. They therefore cannot be used to directly access information 
in particle data tables. The \texttt{KC} codes range between 1 and 500, 
and give the index to the particle data arrays. Each \texttt{KF} code 
is now one-to-one associated with a \texttt{KC} code; the only 
ambiguity is that \texttt{KC} does not distinguish antiparticles
from particles. Whereas \texttt{KF} codes below 100 still obey 
\texttt{KF = KC}, the mapping of codes above 100 is completely changed. 
It is no longer hard-coded in \texttt{PYCOMP}, but defined by the fourth 
component of the \texttt{KCHG} array. Therefore it can be changed or 
expanded during the course of a run, either by \texttt{PYUPDA} calls or 
by direct user intervention.

\subsection{New Options}

A large amount of new options have been added, related to almost all the 
physics changes above and more, and we here only mention some of the more 
significant ones.

The inclusion of \textsc{Susy} processes means that all the \textsc{SPythia}
\texttt{PYMSSM} commonblock switches and parameters are inherited.
New parameters are added also for other new physics scenarios, such
as technicolor and doubly-charged Higgses. 

The extensions to the physics of virtual photons, outlined in 
section~\ref{sec:photon}, has resulted in two sets of new 
possibilities. One is in the description of the virtual-photon flux, 
where new \texttt{CKIN} switches has been introduced, e.g. to set 
the range of photon $x$ and $Q^2$. This is available when 
\texttt{PYINIT} is called with \texttt{'gamma/lepton'} as beam 
or target, to denote that the photon flux inside the lepton has 
to be considered as a new administrative layer, also documented 
in the event record. The other is the new physics machinery. Here 
the main switch is \texttt{MSTP(14)} that sets the assumed nature 
of the photon or photons, e.g. `a direct photon from the left
collides with a VMD one from the right'. The default is the most
general mixture, meaning 4 components for $\gast\p$ and 13 for
$\gast\gast$. This is the relevant approach for studies of QCD 
processes. There is no corresponding automatic mixing machinery
for other processes, so then the relevant contributing components
have to be handled separately and added afterwards. Further options
are available for several of the components, e.g. the DIS 
process dampening in the $Q^2 \to 0$ limit, the relative
normalization of the GVMD spectrum, the scale choice for
parton distributions, and the possibility to add the effects of
a longitudinal resolved contribution.  

The matrix-element options for $\e^+\e^- \to \gammaZ \to$
2, 3 or 4 partons have previously only been available via the
\texttt{LUEEVT}/\texttt{PYEEVT} routine, that suffers from problems
of its own in having a rather old-fashioned machinery for QED 
initial-state radiation and electroweak parameters. Now the QCD 
matrix-element description is accessible as an option to the shower 
default for $\e^+\e^-$ events generated with subprocess 1 of the 
standard \textsc{Pythia} machinery. 

\subsection{Interfaces}

While \textsc{Pythia} contains an extensive library of subprocesses,
it is far from up to all the requirements of the experimental
community. Both further processes and a more detailed treatment 
of the existing ones is required at times. In particular, it is not
uncommon with a generator dedicated to one specific process, where 
also higher-order electroweak corrections, absent in \textsc{Pythia},
have been included in the cross section. None of these programs are
geared to handle the QCD aspects of parton showers and hadronization,
however, so it makes sense to combine the individual strengths.

A generic facility to include external processes exists since long in 
\textsc{Pythia}. Here one can feed in partonic configurations from
an external generator, together with some basic information on 
colour flow and which partons are allowed to radiate, and let 
\textsc{Pythia} construct a complete event based on this information.
For the simple configurations encountered in $\e^+\e^-$ annihilation
events, this would often be overkill, since neither the initial-state 
QCD radiation nor beam-remnant treatment of the generic (hadronic) 
collision is present. 

Based on the concepts presented in the LEP2 workshop \cite{LEP2QCD},
a few simpler alternatives are therefore now provided for 
this kind of tasks:
\begin{Itemize}
\item \texttt{CALL PY2FRM(IRAD,ITAU,ICOM)} allows a parton shower to 
develop and partons to hadronize from a given two-fermion starting 
point. \texttt{IRAD} sets whether quarks are allowed also to radiate 
photons or not, \texttt{ITAU} whether $\tau$ leptons should be decayed 
or not, and \texttt{ICOM} whether the input and output event record is 
\texttt{HEPEVT} or \texttt{PYJETS}. An arbitrary number of photons 
(e.g. from initial-state radiation) may also be stored with the input.  
\item \texttt{CALL PY4FRM(ATOTSQ,A1SQ,A2SQ,ISTRAT,IRAD,ITAU,ICOM)}
allows parton showers to develop and partons to hadronize from a 
given four-fermion starting point.  The extra parameters can be used
to select between the two colour pairings allowed for a 
$\q_1\qbar_2\q_3\qbar_4$ state, according to some different strategies
when interference terms do not allow unique probabilities to be found.
\item \texttt{CALL PY6FRM(P12,P13,P21,P23,P31,P32,PTOP,IRAD,ITAU,ICOM)}
allows parton showers to develop and partons to hadronize from a 
given six-fermion starting point. The \texttt{Pij} parameters give
the relative probabilities for the six colour pairings allowed
for a six-quark state, and \texttt{PTOP} the probability that the
event originates from a $\t\tbar$ pair (in which case the shower
handling has to be different than e.g. in a $\Z^0\W^+\W^-$ event).         
\end{Itemize}

The above routines are not set up to handle QCD four-jet events, i.e.
events of the types $\q\qbar\g\g$ and $\q\qbar\q'\qbar'$, with 
$\q'\qbar'$ coming from a gluon branching. Such events are generated in 
normal parton showers, but not necessarily at the right rate (a problem 
that may be especially interesting for massive quarks like $\b$). Therefore 
one would like to start a QCD parton shower from a given four-parton 
configuration. Some time ago, a machinery was developed to handle 
this kind of occurences \cite{match4jet}. This approach has now been 
adapted to the current \textsc{Pythia} version, in a somewhat modified 
form. In it, an imagined shower history of two branchings is 
(re)constructed from the four-parton state, according to relative 
probabilities derived in the shower language. Thereafter a normal shower 
is allowed to develop, with branchings chosen at random except for these 
two predetermined ones. The routine \texttt{CALL PY4JET(PMAX,IRAD,ICOM)} 
takes an original four-parton configuration stored in \texttt{HEPEVT} or 
\texttt{PYJETS} and lets a shower develop as described above. 
\texttt{PMAX} can be used to set the maximum virtuality of those parts 
of the shower not given from the parton configuration itself, either to 
a fixed value or to the lowest virtuality of the reconstructed shower.

\subsection{Utilities}

The clustering algorithm \texttt{PYCLUS} has been extended also to
accept the Durham distance measure \cite{Durhamclus} as an alternative.
This is $\pT$-based, like the original \texttt{LUCLUS} distance
measure, but differs in the details. 

The \texttt{GBOOK} histogramming package was written in 1979
as a lightweight substitute for \texttt{HBOOK}
\cite{HBOOK} before that program was available in \textsc{Fortran}~77.
The one-dimensional histogram part now appears in the standard
distribution, in order to make the sample runs offered on the web
a bit more realistic. The main routines are:
\begin{Itemize}
\item \texttt{CALL PYBOOK(ID,TITLE,NX,XL,XU)} to book a one-dimensional
histogram with integer identifier \texttt{ID} (in the range $1 - 1000$), 
character title \texttt{TITLE} and \texttt{NX} bins stretching from 
\texttt{XL} to \texttt{XU}.
\item \texttt{CALL PYFILL(ID,X,W)} to fill histogram \texttt{ID}
at position \texttt{X} with weight \texttt{W}.
\item \texttt{CALL PYFACT(ID,F)} to rescale the contents of histogram 
\texttt{ID} by a factor \texttt{F}.
\item \texttt{CALL PYOPER(ID1,OPER,ID2,ID3,F1,F2)} to perform operations
on several histograms, such as adding or dividing them by each other.
\item \texttt{CALL PYDUMP(MDUMP,LFN,NHI,IHI)} to dump histogram contents
to a file from which they could be read in for plotting in another 
program.
\item \texttt{CALL PYHIST} to print all histograms in a simple 
line-printer mode, and thereafter reset histogram contents.
\end{Itemize}
A commonblock of dimension 20000 is used to store the histograms;
this size may need to be expanded if many histograms are to be booked.

The \texttt{PYUPDA} routine has been expanded with a new option that
allows a set of particle data to be read in, in tabular form as 
before, as an addition to or partial replacement of the existing 
particle data. 

\section{Summary and Outlook}

We have here given a very brief survey of news in the 
\textsc{Pythia}~6.1 program. A more detailed description of 
physics and programs is available separately \cite{longPyt}. 
Any serious user should turn to this publication, and to the 
original physics papers, for further information. 

The treasure trove for information is the \textsc{Pythia} webpage,\\
\hspace*{\fill}
\texttt{http://www.thep.lu.se/}$\sim$\texttt{torbjorn/Pythia.html}~,
\hspace*{\fill}\\
where one may find the current and previous subversions, with
documentation, sample main programs, links to related programs, etc.

The \textsc{Pythia} program is continuously being developed. 
We are aware of many physics shortcomings, which hopefully will 
be addressed in the future. It is in the nature of a program of 
this kind never to be finished, at least as long as it is of 
importance for the high-energy physics experimental community.  

The main visible change in the future is the transition to C++ 
as the programming language for \textsc{Pythia}~7. Even if much of 
the physics will
be carried over unchanged, none of the existing code will survive.
The structure of the event record and the whole administrative
apparatus is completely different from the current one, in order
to allow a much more general and flexible formulation of the
event generation process. Following the formulation of a strategy
document \cite{pythia7strat}, a first proof-of-concept version
was released recently \cite{pythia7debut}. So far it only
contains one reasonably complete physics module, however, 
namely that of string fragmentation. More realistic versions 
should follow, but it will take a long time to convert all 
important physics components from \textsc{Pythia}~6. The two
versions therefore will coexist for several years, with the
\textsc{Fortran} one used for physics `production' and the C++ 
one for exploration of the object-oriented approach that will
be standard at the LHC.

\subsection*{Acknowledgements}

A large number of persons should be thanked for their contributions.
Bo Andersson and G\"osta Gustafson are the originators of the Lund 
model, and strongly influenced the early development of the programs.
Hans-Uno Bengtsson is the originator of the \textsc{Pythia} program. 
Mats Bengtsson is the original author of the final-state parton-shower 
algorithm. Several pieces of code have been donated by other persons.
Further comments on the programs have been obtained from users too 
numerous to be mentioned here, but who are all gratefully acknowledged. 
To write programs of this size and complexity would be impossible 
without a strong user feedback.

\end{document}